\documentclass[aps,preprint,nofootinbib,floatfix]{revtex4}
\usepackage{color}

\pdfoutput=1
\usepackage{graphicx}
\usepackage[latin1]{inputenc}
\usepackage{amsmath,amssymb}
\usepackage{hyperref}
\usepackage{epstopdf}

\newcommand{\lsim}{\mathrel{\mathop{\kern 0pt \rlap
  {\raise.2ex\hbox{$<$}}}
  \lower.9ex\hbox{\kern-.190em $\sim$}}}
\newcommand{\gsim}{\mathrel{\mathop{\kern 0pt \rlap
  {\raise.2ex\hbox{$>$}}}
  \lower.9ex\hbox{\kern-.190em $\sim$}}}

\newcommand{\gev}{\ensuremath{\,\mathrm{GeV}}}
\newcommand{\tev}{\ensuremath{\,\mathrm{TeV}}}

\def  \bcen   {\begin{center}}
\def  \ecen   {\end{center}}
\def  \beq    {\begin{equation}}
\def  \eeq    {\end{equation}}
\def  \bpm    {\begin{pmatrix}}
\def  \epm    {\end{pmatrix}}
\def  \beqa   {\begin{eqnarray}}
\def  \eeqa   {\end{eqnarray}}
\def  \nn     {\nonumber }
\def\bea{\begin{eqnarray}}
\def\eea{\end{eqnarray}}

\def\ga   {\gamma}

\def\th   {\theta}
\def\la   {\lambda}
\def\La   {\Lambda}

\def\sig   {\sigma}

\def\nn{\nonumber}
\def\lee { \left( }
\def\rii { \right) }
\def\lan   {\langle}
\def\ran   {\rangle}
\def\de {\delta}
\def\De {\Delta}

\def\to {\rightarrow}

\begin{document}

{\small
\begin{flushright}
IPMU15-0203 \; LCTS/2015-41 \; DO-TH 15/17 \\
\end{flushright} }

\title{G2HDM : Gauged Two Higgs Doublet Model}
\author{
Wei-Chih Huang$^{1,2}$, Yue-Lin Sming Tsai$^3$, Tzu-Chiang Yuan$^{4,5}$
}

\affiliation{
\small{
$^1$Department of Physics and Astronomy, University College London, United Kingdom\\
$^2$Fakult\"at f\"ur Physik, Technische Universit\"at Dortmund,
44221 Dortmund, Germany \\
$^3$Kavli IPMU (WPI), University of Tokyo, Kashiwa, Chiba 277-8583, Japan\\
$^4$Institute of Physics, Academia Sinica, Nangang, Taipei 11529, Taiwan\\
$^5$Physics Division, National Center for Theoretical Sciences, Hsinchu, Taiwan
}
}

\date{\today}

\begin{abstract}
A novel model embedding the two Higgs doublets in the popular two Higgs doublet models 
into a doublet of a non-abelian gauge group $SU(2)_H$ is presented.
The Standard Model $SU(2)_L$ right-handed fermion singlets are paired up with new heavy fermions to form $SU(2)_H$ doublets, 
while $SU(2)_L$ left-handed fermion doublets are singlets under $SU(2)_H$. 
Distinctive features of this anomaly-free model are: 
(1) Electroweak symmetry breaking is induced from spontaneous symmetry breaking of $SU(2)_H$ via  its triplet vacuum expectation value; 
(2) One of the Higgs doublet can be inert, with its neutral component being 
a dark matter candidate as protected by the $SU(2)_H$ gauge symmetry instead of a discrete $Z_2$ symmetry 
in the usual case;
(3) Unlike Left-Right Symmetric Models, the complex gauge fields $(W_1^{\prime}\mp i W_2^{\prime})$ (along with other complex scalar fields) associated  with the $SU(2)_H$ do {\it not} carry electric charges, 
while the third component $W^{\prime}_3$ can mix with the hypercharge $U(1)_Y$ gauge field 
and the third component of $SU(2)_L$; 
(4) Absence of tree level flavour changing neutral current is guaranteed by gauge symmetry;
and {\it etc}. 
In this work, we concentrate on the mass spectra of scalar and gauge bosons in the model.
Constraints from previous $Z^\prime$ data at LEP  and the Large Hadron Collider measurements of 
the Standard Model Higgs mass, its partial widths of $\ga\ga$ and $Z\ga$ modes are discussed.

\end{abstract}

\maketitle

\section{Introduction \label{section:1}}

Despite the historical discovery of the 125 GeV scalar resonance at the LHC~\cite{Aad:2012tfa, Chatrchyan:2012xdj}
completes the last building block of the Standard Model~(SM), many important questions are still lingering in our minds.
Is this really the Higgs boson predicted by the SM or its imposter? Are there any other scalars hidden somewhere? 
What is the dark matter~(DM) particle? And so on. Extended Higgs sector is commonly used to address various theoretical issues 
such as DM. The simplest solution for DM is to add a gauge-singlet scalar, odd under a $Z_2$ symmetry~\cite{Silveira:1985rk,McDonald:1993ex,Burgess:2000yq}. 
In the context of supersymmetric theories like the Minimal Supersymmetric Standard Model (MSSM) proposed to solve the gauge hierarchy problem,  an additional second Higgs doublet is mandatory due to the requirement of anomaly cancelation and supersymmetry~(SUSY) chiral structure. In the inert two Higgs doublet model~(IHDM)~\cite{Deshpande:1977rw}, since the second Higgs doublet is odd under a $Z_2$ symmetry its neutral component can be a DM candidate~\cite{Ma:2006km,Barbieri:2006dq,LopezHonorez:2006gr,Arhrib:2013ela}.
Moreover the observed baryon asymmetry cannot be accounted for in the SM 
since the CKM phase in the quark sector is too small to generate sufficient baryon asymmetry~\cite{Gavela:1993ts,Huet:1994jb,Gavela:1994dt}, and the Higgs potential cannot
achieve the {\it strong} first-order electroweak phase transition unless the SM Higgs boson 
is lighter than $70$ GeV~\cite{Bochkarev:1987wf,Kajantie:1995kf}.
A general two Higgs doublet model (2HDM) contains additional $CP$-violation source~\cite{Bochkarev:1990fx,Bochkarev:1990gb,Turok:1990zg,Cohen:1991iu,Nelson:1991ab} in the scalar sector and hence it may 
circumvent the above shortcomings of the SM for the baryon asymmetry.

The complication associated with a general 2HDM \cite{Branco:2011iw} 
stems from the fact that there exist many terms in the Higgs potential allowed by the SM gauge symmetry,  including various mixing terms between two Higgs doublets $H_1$ and $H_2$.
In the case of both doublets develop vacuum expectation values (vevs), the observed 125 Higgs boson in general would be a linear combination of three neutral scalars~(or of two $CP$-even neutral scalars if the Higgs potential preserves $CP$ symmetry), resulting in flavour-changing neutral current~(FCNC) at tree-level which is tightly constrained  by experiments.  
This Higgs mixing effects lead to changes on the Higgs decay branching ratios into SM fermions which must be confronted by the
Large Hadron Collider (LHC) data.

One can reduce complexity in the 2HDM Higgs potential by imposing certain symmetry. 
The popular choice is  a discrete symmetry, such as $Z_2$ on the second Higgs doublet in IHDM 
or 2HDM type-I~\cite{Haber:1978jt,Hall:1981bc}, type-II~\cite{Donoghue:1978cj,Hall:1981bc}, 
type-X and type-Y~\cite{Barger:1989fj,Akeroyd:1994ga,Akeroyd:1996he} 
where some of SM fermions are also odd under $Z_2$ unlike IHDM. The other choice is a continuous symmetry, 
such as a local $U(1)$ symmetry discussed in ~\cite{Ko:2012hd,Ko:2013zsa,Ko:2014uka,Ko:2015fxa}.
The FCNC constraints can be avoided by satisfying the alignment condition of the Yukawa couplings~\cite{Pich:2009sp} to eliminating dangerous tree-level contributions although it is not radiatively stable~\cite{Ferreira:2010xe}. Alternatively, the aforementioned $Z_2$ symmetry can be used to evade FCNC at tree-level since SM fermions of the same quantum number only couple to one of two Higgs doublets~\cite{Glashow:1976nt,Paschos:1976ay}.  
Moreover, the Higgs decay branching ratios remain intact in the IHDM since $H_1$ is the only doublet which obtains the vacuum
expectation value~(vev), or one can  
simply make the second Higgs doublet $H_2$ much heavier than the SM one such that $H_2$ essentially decouples from the theory.       
All in all, the IHDM has many merits, including accommodating DM, avoiding stringent collider constraints and having a simpler scalar potential. The required $Z_2$ symmetry, however, is just imposed by hand without justification. The lack of  explanation prompts us to come up with a novel 2HDM, which has the same merits of IHDM with an simpler two-doublet Higgs potential but naturally achieve that
$H_2$ does not obtain a vev, which is reinforced in IHDM by the artificial $Z_2$ symmetry.   

In this work, we propose a 2HDM with additional $SU(2)_H \times U(1)_X$ gauge symmetry, where $H_1$ 
(identified as the SM Higgs doublet) and $H_2$ form an $SU(2)_H$ doublet such that the two-doublet potential itself is as simple as the SM Higgs potential with just a quadratic mass term plus a quartic term.  The price to pay is to introduce additional scalars: one $SU(2)_H$ triplet and one $SU(2)_H$ doublet (which are all singlets under the SM gauge groups) with their vevs providing masses 
to the new gauge bosons. 
At the same time, the vev of the triplet induces the SM Higgs vev, breaking $SU(2)_L\times U(1)_Y$ down to $U(1)_Q$, while $H_2$
do not develop any vev and the neutral component of $H_2$ could be a DM candidate, whose stability is protected by the $SU(2)_H$ gauge symmetry and Lorentz invariance.
In order to write down $SU(2)_H\times U(1)_X$ invariant Yukawa couplings, we introduce heavy $SU(2)_L$ singlet Dirac fermions, the right-handed component of which is paired up with the SM right-handed fermions to comprise $SU(2)_H$ doublets.
The masses of the heavy fermions come from the vev of the $SU(2)_H$ doublet.  
In this setup, the model is anomaly-free with respect to all gauge groups. In what follows, 
we will abbreviate our model as G2HDM, the acronym of gauged 2 Higgs doublet model.

We here concentrate on the scalar and additional gauge mass spectra of G2HDM 
and various collider constraints. DM phenomenology will be addressed in a separate publication. 
As stated before, the neutral component of $H_2$ or any neutral heavy fermion
inside an $SU(2)_H$ doublet can potentially play a role of stable DM due to the $SU(2)_H$ gauge symmetry without resorting
to an ad-hoc $Z_2$ symmetry. It is worthwhile to point out that the way of embedding $H_1$ and $H_2$ into the $SU(2)_H$
doublet is similar to the Higgs bi-doublet in the Left-Right Symmetric Model~(LRSM)~\cite{Mohapatra:1974hk,Mohapatra:1974gc,Senjanovic:1975rk,Mohapatra:1979ia,Mohapatra:1980yp} based on the gauge group $SU(2)_L\times SU(2)_R \times U(1)_{B-L}$, where $SU(2)_L$ gauge bosons connect fields within $H_1$ or $H_2^*$,
whereas the heavy $SU(2)_R$ gauge bosons transform $H_1$ into $H_2^*$.
The main differences between G2HDM and LRSM are: 
\begin{itemize}
\item
The charge assignment on the $SU(2)_H$ Higgs doublet is charged
under $U(1)_Y \times U(1)_X$ while the bi-doublet in LRSM is neutral under $U(1)_{B-L}$.  Thus the corresponding 
Higgs potential are much simpler in G2HDM; 
\item
All $SU(2)_H$ gauge bosons are electrically neutral whereas the $W^{\pm}_R$ of $SU(2)_R$ carry electric charge of one unit;
\item
The SM right-handed fermions such as $u_R$ and $d_R$ do not form a doublet under $SU(2)_H$ unlike
in LRSM where they form a $SU(2)_R$ doublet, leading to very different phenomenology.
\end{itemize}
On the other hand, this model is also distinctive from the twin Higgs model~\cite{Chacko:2005pe,Chacko:2005un} where $H_1$ and $H_2$ are charged under two different gauge groups $SU(2)_A$~(identified as the SM gauge group $SU(2)_L$) and $SU(2)_B$ respectively, and the mirror symmetry on $SU(2)_A$ and $SU(2)_B$, {\it i.e.}, $g_A=g_B$, can be used to cancel quadratic divergence of radiative corrections to the SM Higgs mass.  Solving the little hierarchy problem is the main purpose of twin Higgs model while G2HDM focuses on getting an inert Higgs doublet as the DM candidate without imposing an ad-hoc $Z_2$ symmetry.
Finally, embedding two Higgs doublets into a doublet of a non-abelian gauge symmetry with electrically neutral bosons has been proposed in Refs.~\cite{DiazCruz:2010dc,Bhattacharya:2011tr,Fraser:2014yga}, where the non-abelian gauge symmetry is called $SU(2)_N$ instead of $SU(2)_H$.
Due to the $E_6$ origin, those models have, nonetheless, quite different particle contents for both fermions and scalars as well as
varying embedding of SM fermions into $SU(2)_N$, resulting in distinct phenomenology from our model. 

This paper is organized as follows. 
In Section~\ref{section:model}, we specify the model.
We discuss the complete Higgs potential, Yukawa couplings for SM fermions as well as new heavy fermions and anomaly cancellation.
In Section~\ref{section:spon}, we study spontaneous symmetry breaking conditions (Section~\ref{section:spon_spon}),
and analyze the scalar boson mass spectrum (Section~\ref{section:scalar_mass}) and the extra gauge boson mass spectrum 
(Section~\ref{section:gauge_mass}).
We discuss some phenomenology in Section~\ref{section:higgs_physics}, including numerical solutions for the scalar and gauge bosons masses, $Z'$ constraints, SM Higgs  decays into the $\gamma \gamma$ and $\gamma Z$, and stability of DM candidate. 
Finally, we summarize and conclude in Section~\ref{section:conclusion}. Useful formulas are relegated to two appendixes.

\section{G2HDM Set Up \label{section:model}}

In this section, we first set up the G2HDM by specifying its particle content and write down the Higgs potential, 
including the two Higgs doublets $H_1$ and $H_2$,
an $SU(2)_H$ triplet $\De_H$ and an $SU(2)_H$ doublet $\Phi_H$, where $\De_H$ and $\Phi_H$ are singlets
under the SM gauge group. Second, we spell out the fermion sector, requiring the Yukawa couplings to obey the
$SU(2)_H\times U(1)_X$ symmetry. The abelian $U(1)_X$ factor can be treated as either 
local or global symmetry as will be elaborated  further in later section. 
There are different ways of introducing new heavy fermions but we choose a
simplest realization: the heavy fermions together with the SM right-handed fermions comprise $SU(2)_H$ doublets,
while the SM left-handed doublets are singlets under $SU(2)_H$. 
We note that heavy right-handed neutrinos paired up with 
a mirror charged leptons forming $SU(2)_L$ doublets was suggested before in \cite{Hung:2006ap}.
The matter content of the G2HDM is summarized in Table~\ref{tab:quantumnos}.  
Third, we demonstrate the model is anomaly free.

\begin{table}[htp!]
\begin{tabular}{|c|c|c|c|c|c|}
\hline
Matter Fields & $SU(3)_C$ & $SU(2)_L$ & $SU(2)_H$ & $U(1)_Y$ & $U(1)_X$ \\
\hline \hline
$Q_L=\left( u_L \;\; d_L \right)^T$ & 3 & 2 & 1 & 1/6 & 0\\
$U_R=\left( u_R \;\; u^H_R \right)^T$ & 3 & 1 & 2 & 2/3 & $1$ \\
$D_R=\left( d^H_R \;\; d_R \right)^T$ & 3 & 1 & 2 & $-1/3$ & $-1$ \\
\hline
$L_L=\left( \nu_L \;\; e_L \right)^T$ & 1 & 2 & 1 & $-1/2$ & 0 \\
$N_R=\left( \nu_R \;\; \nu^H_R \right)^T$ & 1 & 1 & 2 & 0 & $1$ \\
$E_R=\left( e^H_R \;\; e_R \right)^T$ & 1 & 1 & 2 &  $-1$  &  $-1$ \\
\hline
$\chi_u$ & 3 & 1 & 1 & 2/3 & 0 \\
$\chi_d$ & 3 & 1 & 1 & $-1/3$ & 0 \\
$\chi_\nu$ & 1 & 1 & 1 & 0 & 0 \\
$\chi_e$ & 1 & 1 & 1 & $-1$ & 0 \\
\hline\hline
$H=\left( H_1 \;\; H_2 \right)^T$ & 1 & 2 & 2 & 1/2 & $1$ \\
$\Delta_H=\left( \begin{array}{cc} \Delta_3/2 & \Delta_p/\sqrt{2}  \\ \Delta_m/\sqrt{2} & - \Delta_3/2 \end{array} \right)$ & 1 & 1 & 3 & 0 & 0 \\
$\Phi_H=\left( \Phi_1 \;\; \Phi_2 \right)^T$ & 1 & 1 & 2 & 0 & $1$ \\
\hline
\end{tabular}
\caption{Matter field contents and their quantum number assignments in G2HDM. 
}
\label{tab:quantumnos}
\end{table}

\subsection{ Higgs Potential}

We have two Higgs doublets, $H_1$ and $H_2$ where $H_1$ is identified as the SM Higgs doublet and $H_2$~(with the same hypercharge $Y=1/2$ as $H_1$)
is the additional $SU(2)_L$ doublet.
In addition to the SM gauge groups, we introduce additional groups, $SU(2)_H\times U(1)_X$ under which $H_1$ and $H_2$ transform as a doublet, $H=(H_1\; H_2)^T$ with $U(1)_X$ charge $X(H)=1$. 

With additional $SU(2)_H$ triplet and doublet, $\De_H$ and $\Phi_H$, which are {\it singlets} under $SU(2)_L$, the Higgs potential invariant under both
$SU(2)_L\times U(1)_Y$ and $SU(2)_H \times U(1)_X$
reads\footnote{Here, we consider renormalizable terms only. In addition, $SU(2)_L$ multiplication is implicit and suppressed.}
\begin{align}
V \left( H , \Delta_H, \Phi_H \right) = V (H) + V (\Phi_H ) + V ( \De_H ) + V_{\rm mix} \left( H , \Delta_H, \Phi_H \right) \; , 
\label{eq:higgs_pot} 
\end{align}
with
\begin{align}
\label{VH1H2}
V (H) =& \; \mu^2_H H^\dag H + \la_H \lee H^\dag H \rii^2 \; ,  \nn \\    
=& \; \mu^2_H   \lee H^\dag_1 H_1 + H^\dag_2 H_2 \rii  +  \la_H \lee H^\dag_1 H_1 + H^\dag_2 H_2  \rii^2  \; , 
\end{align}
which contains just two terms (1 mass term and 1 quartic term) 
as compared to 8 terms (3 mass terms and 5 quartic terms) in general 2HDM \cite{Branco:2011iw};
\begin{align}
\label{VPhi}
V ( \Phi_H ) =& \;  \mu^2_{\Phi}   \Phi_H^\dag \Phi_H  + \la_\Phi \lee \Phi_H^\dag \Phi_H  \rii^2  \; , \nn \\
 =& \;  \mu^2_{\Phi} \lee \Phi^*_1\Phi_1 + \Phi^*_2\Phi_2 \rii 
 +  \la_\Phi \lee \Phi^*_1\Phi_1 + \Phi^*_2\Phi_2 \rii^2 \; , \\
 \label{VDelta}
V ( \De_H ) =& \; - \mu^2_{\De} {\rm Tr} \lee \De_H^\dag \De_H  \rii  \;  + \la_\De \lee {\rm Tr} \lee \De_H^\dag \De_H  \rii \rii^2 \; , \nn \\
= & \; - \mu^2_{\De} \lee \frac{1}{2} \De^2_3 + \De_p \De_m  \rii +  \la_{\De} \lee \frac{1}{2} \De^2_3 + \De_p \De_m  \rii^2 \; , 
\end{align}
and finally the mixed term
\begin{align}
\label{VMix}
V_{\rm{mix}} \left( H , \Delta_H, \Phi_H \right) = 
& \; + M_{H\De}  \lee H^\dag \De_H H \rii -  M_{\Phi\De}  \lee \Phi_H^\dag \De_H \Phi_H \rii  \nn \\
& \;  +  \la_{H\De} \lee H^\dag H  \rii    {\rm Tr} \lee \De_H^\dag \De_H  \rii  
 + \la_{H\Phi} \lee H^\dag H  \rii  \lee \Phi_H^\dag \Phi_H \rii  \nn\\
& \; + \la_{\Phi\De} \lee \Phi_H^\dag \Phi_H \rii {\rm Tr} \lee \De_H^\dag \De_H \rii  \; , \nn \\
= & \; + M_{H\De} \lee \frac{1}{\sqrt{2}}H^\dag_1 H_2 \De_p  
+  \frac{1}{2} H^\dag_1 H_1\De_3 + \frac{1}{\sqrt{2}}  H^\dag_2 H_1 \De_m  
- \frac{1}{2} H^\dag_2 H_2 \De_3   \rii   \nn \\
& \; - M_{\Phi\De} \lee  \frac{1}{\sqrt{2}} \Phi^*_1 \Phi_2 \De_p  
+  \frac{1}{2} \Phi^*_1 \Phi_1\De_3 + \frac{1}{\sqrt{2}} \Phi^*_2 \Phi_1 \De_m  
- \frac{1}{2} \Phi^*_2 \Phi_2 \De_3   \rii  \nn \\
& \; + \la_{H\De} \lee H^\dag_1 H_1 + H^\dag_2 H_2 \rii   \lee \frac{1}{2} \De^2_3 + \De_p \De_m  \rii \nn\\
& \; +  \la_{H\Phi} \lee H^\dag_1 H_1 + H^\dag_2 H_2 \rii  \lee \Phi^*_1\Phi_1 + \Phi^*_2\Phi_2 \rii \nn\\
& \; + \la_{\Phi\De}  
  \lee  \Phi^*_1\Phi_1 + \Phi^*_2\Phi_2 \rii  \lee \frac{1}{2} \De^2_3 + \De_p \De_m  \rii \; , 
\end{align}
where
 \begin{align}
\De_H=
  \begin{pmatrix}
    \De_3/2   &  \De_p / \sqrt{2}  \\
    \De_m / \sqrt{2} & - \De_3/2   \\
  \end{pmatrix} \; {\rm with}
  \;\; \Delta_m = \left( \Delta_p \right)^* \; {\rm and} \; \left( \Delta_3 \right)^* = \Delta_3 \;    ,
 \end{align}
 and $\Phi_H=\left( \Phi_1 \;\; \Phi_2 \right)^T$.

At this point we would like to make some general comments for the above potential 
before performing the minimization of it to achieve spontaneous symmetry breaking (see next Section).
\begin{itemize}

\item
$U(1)_X$ is introduced to simplify the Higgs potential $V \left( H , \Delta_H, \Phi_H \right) $ in Eq.~\eqref{eq:higgs_pot}.
For example, a term $ \Phi^T_H \De_H \Phi_H $ obeying the $SU(2)_H$ symmetry would be allowed in the absence of
$U(1)_X$. Note that as far as the scalar potential is concerned, treating $U(1)_X$ as a global symmetry is sufficient to 
kill this and other unwanted terms.

\item
In Eq.~\eqref{VDelta}, if $-\mu^2_\De <0$, $SU(2)_H$ is spontaneously broken 
by the vev $\lan \De_3 \ran =  - v_\De \neq 0$ with $\lan \De_{p,m} \ran=0$ by applying an $SU(2)_H$ rotation. 

\item
The quadratic terms for $H_1$ and $H_2$ have the following coefficients
\begin{equation}
\mu^2_H \mp \frac{1}{2} M_{H\De} \cdot v_\De + \frac{1}{2} \lambda_{H \De} \cdot v_\De^2 
+  \frac{1}{2} \lambda_{H \Phi} \cdot v_\Phi^2 \; ,
\end{equation} 
respectively.
Thus even with a positive $\mu_H^2$, $H_1$ can still develop a vev $(0 \; v/\sqrt 2)^T$ breaking $SU(2)_L$ provided that the second term is dominant, while $H_2$ remains zero vev.
Electroweak symmetry breaking is triggered by the $SU(2)_H$ breaking.
Since the doublet $H_2$ does not obtain a vev,  its lowest mass component can be 
potentially a DM candidate whose stability is protected by the gauge group $SU(2)_H$. 

\item
Similarly, the quadratic terms for two fields $\Phi_1$ and $\Phi_2$ have the coefficients
\begin{equation}
\mu^2_\Phi \pm \frac{1}{2} M_{\Phi\De} \cdot v_\De + \frac{1}{2} \lambda_{\Phi \De} \cdot v_\De^2 
+  \frac{1}{2} \lambda_{H \Phi} \cdot v^2 \; ,
\end{equation}
respectively. The field $\Phi_2$ may acquire nontrivial vev  and $\lan \Phi_1 \ran =0$ with the help of a large second term.

\end{itemize}

\subsection{Yukawa Couplings}

We start from the quark sector. Setting the quark $SU(2)_L$ doublet, $Q_L$, to be an $SU(2)_H$ singlet and 
including additional $SU(2)_L$ singlets $u^H_R$ and $d^H_R$ which together with the SM right-handed quarks $u_R$ and $d_R$, respectively, to form
$SU(2)_H$ doublets, i.e., $U_R^T = (u_R \;\, u^H_R)_{2/3}$ and $D_R^T = (d^H_R \;\, d_R)_{-1/3}$,
where the subscript represents hypercharge, we have~\footnote{$A \cdot B$ is defined as 
$\epsilon_{ij}A^i B^j$ where $A$ and $B$ are two 2-dimensional spinor representations of $SU(2)_H$.}
\begin{align}
\mathcal{L}_{\rm Yuk} \supset & \; y_d \bar{Q}_L \lee D_R \cdot H \rii 
+ y_u \bar{Q}_L \lee U_R \cdot \stackrel{\thickapprox}{H}   \rii+ {\rm H.c.} , \nn\\
=& \; y_d \bar{Q}_L \lee d^H_R H_2 - d_R H_1  \rii - y_u \bar{Q}_L \lee u_R \tilde{H}_1 + u^H_R \tilde{H}_2  \rii   + {\rm H.c.},
\label{eq:Yuk_Q}
\end{align}
where $\stackrel{\thickapprox}{H} \equiv ( \tilde{H}_2 \;  - \tilde{H}_1 )^T$ with 
$\tilde H_{1,2} = i \tau_2 H_{1,2}^*$.
After the EW symmetry breaking $\lan H_1\ran \not= 0$, $u$ and $d$ obtain their masses but $u^H_R$ and $d^H_R$ remain massless since $H_2$ does not get a vev. 

To give a mass to the additional species, we employ the $SU(2)_H$ scalar 
doublet $\Phi_H = (\Phi_1 \; \Phi_2)^T$, which is singlet under $SU(2)_L$,
and left-handed $SU(2)_{L,H}$ singlets $\chi_u$ and $\chi_d$ as
\begin{align}
\mathcal{L}_{\rm Yuk} \supset & \;  -  y^\prime_d \overline{\chi}_d \lee D_R \cdot \Phi_H \rii 
+ y'_u \overline{\chi}_u \lee U_R \cdot \tilde{\Phi}_H \rii  + {\rm H.c.} , \nn\\
=& \;  -  y'_d \overline{\chi}_d \lee d^H_R \Phi_2 - d_R \Phi_1  \rii 
- y'_u \overline{\chi}_u \lee u_R \Phi^*_1 + u^H_R \Phi^*_2  \rii   + {\rm H.c.}, 
\label{eq:Yuk_Q1} 
\end{align}
where $\Phi$ has $Y=0$, $Y(\chi_u)=Y(U_R)=2/3$ and $Y(\chi_d)=Y(D_R)=-1/3$ with
$\tilde{\Phi}_H=( \Phi^*_2 \; - \Phi^*_1 )^T$. With $\lan \Phi_2 \ran= v_{\Phi}/ \sqrt{2}$, $u^H_R~(\chi_u)$ and $d^H_R~(\chi_d)$ obtain masses
$y'_u v_{\Phi}/ \sqrt{2}$ and $y'_d v_{\Phi}/ \sqrt{2}$, respectively.
Note that both $v_\De$ and $v_{\Phi}$ contribute the $SU(2)_H$ gauge boson masses.
 
The lepton sector is similar to the quark sector as 
\begin{align}
\mathcal{L}_{\rm Yuk} \supset & \;  y_e \bar{L}_L \lee E_R \cdot H \rii 
+ y_\nu {\bar L}_L \lee N_R \cdot \stackrel{\thickapprox}{H} \rii 
 -  y'_e \overline{\chi}_e \lee E_R \cdot \Phi_H \rii   
+ y'_\nu \overline{\chi}_\nu \lee N_R \cdot \tilde \Phi_H \rii   
+ {\rm H.c.} , \nn\\
=& \; y_e \bar{L}_L \lee e^H_R H_2 - e_R H_1 \rii 
- y_\nu \bar{L}_L \lee \nu_R \tilde{H_1} + \nu^H_R \tilde{H_2} \rii \nn\\
& \; -  y'_e \overline{\chi}_e \lee e^H_R \Phi_2 - e_R \Phi_1  \rii   
- y'_\nu \overline{\chi}_\nu \lee \nu_R \Phi^*_1 + \nu^H_R \Phi^*_2 \rii
+ {\rm H.c.},
\label{eq:Yuk_L}
\end{align}
where $E_R^T = (e^H_R  \; e_R)_{-1}$, $N_R^T = (\nu_R  \;  \nu^H_R)_{0}$ in which $\nu_R$ and $\nu^{H}_R$
are the right-handed neutrino and its $SU(2)_H$ partner respectively, 
while $\chi_e$ and $\chi_\nu$ are $SU(2)_{L,H}$ singlets with $Y(\chi_e)=-1$ and $Y(\chi_\nu)=0$ respectively.
Notice that neutrinos are purely Dirac in this setup, {\it i.e.},
$\nu_R$ paired up with $\nu_L$ having Dirac mass $M^{\nu}_D = y_\nu v /\sqrt 2$, 
while $\nu^H_R$ paired up with $\chi_\nu$ having Dirac mass $M^{\nu^H}_D = y^\prime_\nu v_{\Phi} /\sqrt 2$.
As a result, the lepton number is conserved, implying vanishing neutrinoless double beta decay. 
In order to generate the observed neutrino masses of order sub-eV, 
the Yukawa couplings for $\nu_L$ and $\nu_R$
are extremely small~($\sim 10^{-11}$) even compared to the electron Yukawa coupling. 
The smallness can arise from, for example, the small overlap among wavefunctions
along a warped extra dimension~\cite{Randall:1999ee,Randall:1999vf}.  

Alternatively, it may be desirable for the neutrinos to have a Majorana mass term which can be easily incorporated by 
introducing a $SU(2)_H$ scalar triplet $\De_N$ with $X(\De_N)=-2$. Then a renormalizable term 
$g_N\overline{N^c_R} \De_N N_R$ with a large $\lan \De_N \ran \neq 0$ will
break lepton number and provide large Majorana masses $M_N = g_N \langle \De_N \rangle$ 
to $\nu_R$s~(and also $\nu^H_R$s). 
Sub-eV masses for the $\nu_L$s can be realized 
via the type-I seesaw mechanism which allows one large mass of order $M_N$ and one small mass of order 
$(M^\nu_D)^2/M_N$. For $M^\nu_D \sim y_\nu v$ and $v \sim$ 246 GeV, sub-eV neutrino masses can be achieved provided that
$y_\nu \sim 1.28 \times 10^{-7} \sqrt{M_N/{\rm GeV}}$.

We note that only one $SU(2)_L$ doublet $H_1$ couples to two SM fermion fields in the above Yukawa couplings. 
The other doublet $H_2$ couples to one SM fermion 
and one non-SM fermion, while the $SU(2)_H$ doublet $\Phi_H$ couples to at least one non-SM fermion. 
As a consequence, there is no flavour changing decays from the SM Higgs 
in this model. This is in contrast with the 2HDM where a discrete $Z_2$ symmetry needed to be imposed to 
forbid flavour changing Higgs decays at tree level. Thus, as long as $H_2$ does not develop a vev in the parameter space, 
it is practically an inert Higgs, protected by a local gauge symmetry instead of a discrete one!

\subsection{Anomaly Cancellation}
 
We here demonstrate the aforementioned setup is anomaly-free with respect to both the SM and additional gauge groups. 
The anomaly cancellation for the SM gauge groups $SU(3)_C \times SU(2)_L \times U(1)_Y$ is guaranteed since addition heavy particles of the same hypercharge
form Dirac pairs.
Therefore, contributions of the left-handed currents from $\chi_u$, $\chi_d$, $\chi_\nu$ and $\chi_e$ cancel those of right-handed ones from
$u^H_R$, $d^H_R$, $\nu^H_R$ and $e^H_R$ respectively.

Regarding the new gauge group $SU(2)_H$, the only nontrivial anomaly needed to be checked is 
$[SU(2)_H]^2 U(1)_Y$ from the doublets $U_R$, $D_R$, $N_R$ and $E_R$ with the following result
\begin{align}
2 {\rm Tr}[T^a \{ T^b , Y \}]  = & 2 \de^{ab} \left( \sum_l Y_l - \sum_r Y_r  \right) =
- 2 \de^{ab}  \sum_r Y_r  \nonumber \\
 = & - 2 \de^{ab}  \left( 3 \cdot 2 \cdot Y(U_R) +   3 \cdot 2 \cdot Y(D_R) +  2 \cdot Y(N_R) + 2 \cdot Y(E_R) \right)
\end{align}
where $3$ comes from the $SU(3)_C$ color factor and $2$ from $2$ components in an $SU(2)_H$ doublet.
With the quantum number assignment for the various fields listed in Table~\ref{tab:quantumnos}, one can check that
this anomaly coefficient vanishes for each generation.

In terms of $U(1)_X$, one has to check $[SU(3)_C]^2 U(1)_X$, $[SU(2)_H]^2 U(1)_X$, $[U(1)_X]^3$,
$[U(1)_Y]^2 U(1)_X$ and $[U(1)_X]^2 U(1)_Y$.\footnote{ $[SU(2)_L]^2 U(1)_X$ anomaly 
does not exist since fermions charged under $U(1)_X$ are singlets under $SU(2)_L$.} 
The first three terms are zero due to cancellation between $U_R$ and
$D_R$ and between $E_R$ and $N_R$ with opposite $U(1)_X$ charges.
For $[U(1)_Y]^2 U(1)_X$ and $[U(1)_X]^2 U(1)_Y$, one has respectively 
\begin{align}
 & 2 \cdot \lee 3 \cdot \lee Y(U_R)^2 X(U_R) + Y(D_R)^2 X(D_R)  \rii  +  Y(E_R)^2 X(E_R)  \rii   , \nn \\
 & 2 \cdot \lee 3 \cdot \lee X(U_R)^2 Y(U_R) + X(D_R)^2 Y(D_R)  \rii  +  X(E_R)^2 Y(E_R)  \rii ,  
\end{align}
both of which vanish.

One can also check the perturbative gravitational anomaly~\cite{AlvarezGaume:1983ig} associated with the hypercharge and $U(1)_X$ charge
current couples to two gravitons 
is proportional to the following sum of the hypercharge 
\begin{align}
3 \cdot \left( 2  \cdot  Y(Q_L)+ Y(\chi_u) + Y(\chi_d)  - 2  \cdot Y(U_R) - 2  \cdot Y(D_R) \right) \nonumber \\
+ 2  \cdot Y(L_L) + Y (\chi_\nu) + Y(\chi_e)  - 2  \cdot Y(N_R) - 2  \cdot Y(E_R)  ,
\end{align}
and $U(1)_X$ charge
\begin{align}
X(U_R) + X(D_R) + X(E_R) + X(N_R) ,
\end{align}
which also vanish for each generation.
 
Since there are 4 chiral doublets for $SU(2)_L$ and  also 8 chiral doublets for $SU(2)_H$ for each generation,
the model is also free of the global $SU(2)$ anomaly~\cite{Witten:1982fp} which requires the total number of 
chiral doublets for any local $SU(2)$ must be even.

We end this section by pointing out that one can also introduce $Q^H_L = (u^H_L \; d_L^H )^T$ to pair up with $Q_L$
and $L^H_L = (\nu^H_L \; e_L^H )^T$ to pair up with $L_L$
to form $SU(2)_H$ doublets. Such possibility is also interesting 
and will be discussed elsewhere. 

\section{Spontaneous Symmetry Breaking and Mass Spectra } \label{section:spon}
 
After specifying the model content and fermion mass generation, we now switch to the scalar and gauge boson sector.  We begin by studying the minimization conditions for spontaneous symmetry 
breaking, followed by investigating scalar and gauge boson mass spectra.
 Special attention is paid to mixing effects on both the scalars and gauge bosons. 

\subsection{Spontaneous Symmetry Breaking}\label{section:spon_spon}
 
To facilitate spontaneous symmetry breaking, let us shift the fields as follows
\begin{eqnarray}
H_1 = 
\begin{pmatrix}
G^+ \\ \frac{v + h}{\sqrt 2} + i G^0
\end{pmatrix}
\;\; , \;\; 
\Phi_H = 
\begin{pmatrix}
G_H^p \\ \frac{v_\Phi + \phi_2}{\sqrt 2} + i G_H^0
\end{pmatrix}
\;\; , \;\; 
\Delta_H =
\begin{pmatrix}
\frac{-v_\De + \delta_3}{2} & \frac{1}{\sqrt 2}\De_p \\ 
\frac{1}{\sqrt 2}\De_m & \frac{v_\De - \delta_3}{2}
\end{pmatrix}
\end{eqnarray}
and $H_2=(H_2^+ \; H_2^0)^T$. Here $v$, $v_\Phi$ and $v_\De$ are vevs to be determined
by minimization of the potential; $\Psi_G \equiv \{ G^+, G^3, G^p_H,  G^0_H\}$ are Goldstone bosons, to be
absorbed by the longitudinal components of $W^+$, $W^3$, $W^p$, $W^{\prime 3}$ respectively; 
and $\Psi \equiv \{ h,H_2,\Phi_1,\phi_2, \de_3, \De_p \}$ are the physical fields.

Substituting the vevs in the potential $V$ in Eq.~\eqref{eq:higgs_pot}  leads to
\begin{eqnarray}
\label{Vvevs}
V(v,  v_\De , v_\Phi) & = & 
\frac{1}{4} \left[ 
\lambda_H v^4 + \lambda_\Phi v_\Phi^4 + \lambda_\De v_\De^4 + 2 \left( \mu_H^2 v^2 + \mu_\Phi^2 v_\Phi^2 - \mu_\De^2 v_\De^2 \right) \right. \nonumber \\
&& \left. - \left( M_{H\De} v^2 + M_{\Phi\De} v_\Phi^2 \right) v_\De + \lambda_{H\Phi} v^2 v_\Phi^2 + \lambda_{H\De} v^2 v_\De^2 + \lambda_{\Phi\De} v_\Phi^2 v_\De^2
\right]
\end{eqnarray}
Minimization of the potential in Eq.~\eqref{Vvevs}
leads to the following three equations for the vevs
\begin{eqnarray}
\label{vevv}
v \cdot \left( 2\lambda_H v^2 + 2 \mu_H^2 - M_{H\De} v_\De  + \lambda_{H\Phi} v_\Phi^2 + \lambda_{H\De} v_\De^2 \right)
& = & 0 \; , \\
\label{vevphi}
v_\Phi \cdot \left( 2\lambda_\Phi v_\Phi^2 + 2 \mu_\Phi^2 -  M_{\Phi\De} v_\De + \lambda_{H\Phi} v^2 + \lambda_{\Phi\De} v_\De^2 \right)
& = &  0 \; , \\
\label{vevdelta}
4\lambda_\De v_\De^3 - 4 \mu_\De^2 v_\De - M_{H \De} v^2 - M_{\Phi \De} v_\Phi^2 
+ 2 v_\De \left( \lambda_{H\De} v^2 + \lambda_{\Phi\De} v_\Phi^2 \right) & = & 0 \; .
\end{eqnarray}
Note that one can solve for the non-trivial solutions for 
$v^2$ and $v_\Phi^2$ in terms of $v_\De$ and other parameters using Eqs.~\eqref{vevv}
and \eqref{vevphi}. Substitute these solutions of $v^2$ and $v_\Phi^2$ into Eq.~\eqref{vevdelta} leads to a cubic equation
for $v_\De$ which can be solved analytically (See Appendix~\ref{section:app}).


\subsection{Scalar Mass Spectrum}\label{section:scalar_mass}
 
The scalar boson mass spectrum can be obtained from taking the second derivatives of 
the potential with respect to the various fields and evaluate it at the minimum of the potential.
The mass matrix thus obtained contains three diagonal blocks. The first block is $3 \times  3$.
In the basis of $S=\{h, \delta_3, \phi_2\}$ it is given by
\begin{align}
{\mathcal M}_0^2 =
\begin{pmatrix}
2 \lambda_H v^2 & \frac{v}{2} \left( M_{H\De} - 2 \lambda_{H \De} v_\De \right) & \lambda_{H \Phi} v v_\Phi \\
\frac{v}{2} \left( M_{H\De} - 2 \lambda_{H \De} v_\De \right) & 
\frac{1}{4 v_\De} \left( 8 \lambda_\De v_\De^3 + M_{H\Delta} v^2 + M_{\Phi \De} v_\Phi^2 \right)   &  \frac{ v_\Phi}{2} \left( M_{\Phi\De} - 2 \lambda_{\Phi \De} v_\De \right) \\
\lambda_{H \Phi} v v_\Phi & \frac{ v_\Phi}{2} \left( M_{\Phi\De} - 2 \lambda_{\Phi \De} v_\De \right) & 2 \lambda_\Phi v_\Phi^2
\end{pmatrix} \; .
\label{eq:scalarbosonmassmatrix}
\end{align}
This matrix can be diagonalized by a similar transformation with orthogonal matrix  $O$, which defined as
$ \vert f \rangle_i \equiv O_{i j}  \vert m \rangle_j $ with $i$ and $j$ referring to the flavour and mass eigenstates respectively,  
\begin{equation}
O^T \cdot {\mathcal M}_0^2 \cdot O = {\rm Diag}(m^2_{h_1}, m^2_{h_2}, m^2_{h_3}) \; ,
\end{equation}
where the three eigenvalues are in ascending order.
The lightest eigenvalue $m_{h_1}$ will be identified as the 125 GeV Higgs $h_1$ observed at the LHC 
and the other two $m_{h_2}$ and $m_{h_3}$ are for the heavier Higgses $h_2$ and $h_3$. The physical Higgs $h_i$ is 
a linear combination of  the three components of $S$: $h_i = O_{ji}S_j$. 
Thus the 125 GeV scalar boson could be a mixture of the neutral components of 
$H_1$ and the $SU(2)_H$ doublet $\Phi_H$, 
as well as the real component $\delta_3$ of the $SU(2)_H$ triplet $\Delta_H$.

The SM Higgs $h_1$ tree-level couplings to $f \bar f$, $W^+W^-$, $ZZ$ and $H_2^+H_2^-$ 
pairs, each will be modified by an overall factor of $O_{11}$, 
resulting a reduction by $ | O_{11} |^2 $ on the $h_1$ decay branching ratios
into these channels. On the other hand, as we shall see later, $h_1 \to \gamma\gamma$ and $Z \gamma$ involve 
extra contributions from the $\delta_3$
and $\phi_2$ components, which could lead to either enhancement or suppression with respect to the SM prediction.

The second block is also $3 \times  3$. In the basis of 
$G=\{ G^p_H , \De_p , H^{0*}_2 \}$ it is given by
\begin{align}
{\mathcal M}_0^{\prime 2} =
\begin{pmatrix}
M_{\Phi \De} v_\De & - \frac{1}{2} M_{\Phi \De} v_\Phi & 0 \\
- \frac{1}{2} M_{\Phi \De} v_\Phi &  \frac{1}{4 v_\De} \left( M_{H\De} v^2 + M_{\Phi \De} v_\Phi^2 \right) &  
\frac{1}{2} M_{H \De} v\\
0 & \frac{1}{2} M_{H \De} v & M_{H \De} v_\De
\end{pmatrix} \; .
\label{goldstonemassmatrix}
\end{align}
It is easy to show that Eq.~\eqref{goldstonemassmatrix} has a zero eigenvalue, associated with the 
physical Goldstone boson, which is a mixture of $G^p_H$, $\Delta_p$ and $H^{0*}_2$. 
The other two eigenvalues are the masses of two physical fields $\widetilde \Delta$ and $D$.
They are given by
\begin{eqnarray}
\label{darkmattermass}
M_{{\widetilde \Delta},D} &=&
\frac{1}{8 v_\De} \Bigl\{
M_{H\De} v^2 + 4 \left( M_{H\De}+M_{\Phi\De} \right) v_\De^2 + M_{\Phi\De} v_\Phi^2 
\Bigr.  \\
& \pm & \left.  \left[ 
\left( M_{H \De} \left( v^2 + 4 v_\De^2 \right) + M_{\Phi\De} \left( v_\Phi^2 + 4 v_\De^2 \right) \right)^2
- 16 M_{H\De}M_{\Phi\De} v_\De^2 \left( v^2 + 4 v_\De^2 + v_\Phi^2 \right)
\right]^{\frac{1}{2}}\right\} \; .\nonumber
\end{eqnarray}
$D$ can be a DM candidate in G2HDM. Note that in the parameter space where the quantity inside the square root 
of Eq.~\eqref{darkmattermass} is very small, $\widetilde \Delta$ would be degenerate with $D$. In this case, we need to include
coannihilation processes for relic density calculation.
Moreover, it is possible in our model to have $\nu_R^H$ or $\chi_\nu$~($\nu_R$ either is too light or is not stable since it decays to SM lepton and Higgs) 
to be DM candidate as well.

The final block is $4 \times 4$ diagonal, giving 
\begin{equation}
m^2_{H_2^\pm} = M_{H \De} v_\De \; ,
\end{equation}
for the physical charged Higgs $H_2^\pm$, and 
\begin{equation}
m^2_{G^\pm} = m^2_{G^0} = m^2_{G^0_H} = 0 \; ,
\end{equation}
for the three Goldstone boson fields $G^\pm$, $G^0$ and $G^0_H$.
Note that we have used the minimization conditions Eqs.~\eqref{vevv}, \eqref{vevphi} and \eqref{vevdelta} 
to simplify various matrix elements of the above mass matrices.

Altogether we have 6 Goldstone particles in the scalar mass spectrum, we thus expect to have two massless gauge particles
left over after spontaneous symmetry breaking. One is naturally identified as the photon while the other one could be interpreted as dark photon
$\gamma_D$.

\subsection{$SU(2)_H \times U(1)_X $ Gauge Boson Mass Spectrum } \label{section:gauge_mass}

After investigating the spontaneous symmetry breaking conditions, we now study the mass spectrum of additional gauge bosons.
The gauge kinetic terms for the $\Delta_H$, $\Phi$ and $H$ are 
\begin{align}
\mathcal{L} \supset {\rm Tr} \left[
\lee D^{\prime}_\mu \Delta_H \rii^\dag \lee D^{\prime\mu} \Delta_H  \rii \right]
+ \lee D^{\prime}_\mu \Phi\rii^\dag \lee D^{\prime\mu} \Phi\rii
+  \lee D^{\prime}_\mu H \rii^\dag \lee D^{\prime\mu} H \rii,
\end{align}
with
\begin{align}
D^{\prime}_\mu \Delta_H =  \partial_\mu \Delta_H - i g_H \left[ W^{\prime}_\mu , \Delta_H \right] \;\; ,
\end{align}
\begin{align}
D^{\prime}_\mu \Phi = \lee \partial_\mu - i \frac{g_H}{\sqrt{2}} \lee W^{\prime p}_\mu T^p + W^{\prime m}_\mu T^m \rii - i g_H W^{\prime 3}_\mu T^3  - i  g_X  X_\mu \rii \cdot \Phi \;\; ,
\end{align}
and
\begin{align}
D^{\prime}_\mu H =& \Bigl( D_\mu \cdot 1  - i \frac{g_H}{\sqrt{2}} \lee W^{\prime p}_\mu T^p + W^{\prime m}_\mu T^m \rii - i g_H W^{\prime 3}_\mu T^3  - i  g_X  X_\mu  \Bigl)  \cdot  H \;\; ,
\end{align}
where $D_\mu$ is the $SU(2)_L$ covariant derivative,
acting individually on $H_1$ and $H_2$, $g_H~(g_X)$ is the $SU(2)_H~(U(1)_X)$ gauge coupling constant,
and
\begin{align}
W^\prime_\mu = \sum_{a=1}^3 W^{\prime a} T^a
= \frac{1}{\sqrt{2}} \lee W^{\prime p}_\mu T^p + W^{\prime m}_\mu T^m \rii +W^{\prime 3}_\mu T^3 ,
\end{align}
in which 
$T^a = \tau^a/2$ ($\tau^a$ are the Pauli matrices acting on the $SU(2)_H$ space), 
$W^{\prime\, (p,m)}_\mu=( W^{\prime 1}_\mu \mp i W^{\prime \, 2}_\mu ) /{\sqrt 2}$,
and 
\begin{align}
T^p =  \frac{1}{2} \left( \tau^1 + i \tau^2 \right) =
\begin{pmatrix}
    0   & 1  \\
    0 & 0   \\
  \end{pmatrix}, \; 
T^m =   \frac{1}{2} \left( \tau^1 - i \tau^2 \right) =
\begin{pmatrix}
    0   & 0  \\
    1 & 0   \\
  \end{pmatrix} \;\; . 
\end{align}

The SM charged gauge boson $W^{\pm}$ obtained its mass entirely from $v$, so it is given by
\begin{equation}
M_{W^\pm} = \frac{1}{2} g v \; ,
\end{equation}
same as the SM. 

The $SU(2)_H$ gauge bosons $W^{\prime a}$
and the $U(1)_X$ gauge boson $X$ receive masses from $\lan \De_3 \ran$, $\lan H_1 \ran$
and $\lan \Phi_2 \ran$. The terms contributed from the doublets are similar with that from the standard model.
Since $\De_H$ transforms as a triplet under $SU(2)_H$, i.e., in the adjoint
representation, the contribution to the $W^{\prime a}$ masses arise from the term
\begin{align}
\mathcal{L} \supset g^2_H {\rm Tr} \lee  \left[W^{\prime\mu} , \De_H \right]^\dag
\left[W^\prime_\mu ,\De_H\right] \rii.
\end{align}

All in all, the $W^{\prime (p,m)}$ receives a mass from $\lan \De_3 \ran$, $\lan \Phi_2 \ran$ and $\lan H_1 \ran$
\begin{align}
m^2_{W^{\prime (p,m)}}  = \frac{1}{4} g^2_H \lee v^2 + v^2_\Phi + 4 v^2_\De \rii,  \; 
\end{align}
while gauge bosons $X$ and $W^{\prime 3}$, together with the SM $W^3$ and $U(1)_Y$ gauge boson $B$,
acquire their masses from $\lan \Phi_2 \ran$ and $\lan H_1 \ran$ only but not from $\lan \Delta_H \ran$:
\begin{align}
\label{neutralbosonmasses}
\frac{1}{8} \lee  v^2 \lee 2 g_X X_\mu + g_H W_\mu^{\prime 3} - g W_\mu^{3} + g^\prime B_\mu \rii^2
 + v^2_\Phi \lee -2 g_X X_\mu + g_H W_\mu^{\prime 3} \rii^2     \rii ,
\end{align}
where $g^\prime$ is the SM $U(1)_Y$ gauge coupling.
 
Note that the gauge boson $W^{\prime {\lee p,m \rii }}$ corresponding to the $SU(2)_H$ generators $T^{\pm}$ do {\it not} carry
the SM electric charge and therefore will {\it not} mix with the SM $W^\pm$ bosons while $W^{\prime 3}$ and $X$
{\it do} mix with the SM $W^3$ and $B$ bosons via $\lan H_1 \ran$. In fact, only two of $W^3$, $W^{\prime 3}$, $B$ and $X$
will become massive, by absorbing the imaginary part of $H^0_1$ and $\Phi_2$. To avoid undesired additional massless gauge bosons,
one can introduce extra scalar fields charged under only $SU(2)_H \times U(1)_X$ but not under
 the SM gauge group to give a mass to $W^{\prime 3}$ and $X$, without perturbing the SM gauge boson mass spectrum.
 Another possibility is to involve the Stueckelberg mechanism to give a mass to the $U(1)_X$ gauge boson as done in
 Refs.~\cite{Kors:2004dx,Kors:2005uz,Feldman:2006ce,Cheung:2007ut}. 
 Alternatively, one can set $g_X=0$ to decouple $X$ from the theory or simply treat $U(1)_X$ as a global symmetry,
 after all as mentioned before $U(1)_X$ is introduced to simplify the Higgs potential by forbidding terms like
 $ \Phi^T_H \De_H \Phi_H $, which is allowed under $SU(2)_H$ but not $U(1)_X$.

From Eq.~\eqref{neutralbosonmasses},  one can obtain the following mass matrix for the neutral gauge bosons 
in the basis 
$V^\prime = \left\{ B, W^3, W^{\prime 3}, X \right\}$:
\begin{equation}
\label{M1sq1}
{\mathcal M}_1^2 =     \begin{pmatrix}
\frac{g^{\prime 2} v^2 }{4}  & - \frac{g^{\prime} g \, v^2 }{4}  &  \frac{g^{\prime} g_H v^2 }{4} & \frac{g^\prime g_X v^2}{2}  \\
- \frac{g^{\prime} g \, v^2 }{4} & \frac{ g^2 v^2 }{4} & - \frac{g g_H v^2 }{4} 
& - \frac{ g g_X v^2  }{2} \\
 \frac{g^{\prime} g_H v^2 }{4} & - \frac{g g_H v^2 }{4}   & \frac{g^2_H  \lee v^2 + v^2_\Phi \rii }{4}  & 
 \frac{g_H g_X \lee v^2 - v^2_\Phi \rii }{2} \\ 
 \frac{g^\prime g_X v^2}{2} & - \frac{ g g_X v^2  }{2}
 & \frac{g_H g_X \lee v^2 - v^2_\Phi \rii }{2} & g_X^2 \left( v^2 + v^2_\Phi \right) 
      \end{pmatrix} \; .
\end{equation}
As anticipated, this mass matrix has two zero eigenvalues corresponding to 
$m_\gamma = 0$ and $m_{\gamma_D} = 0$ for the photon and dark photon respectively.
The other two nonvanishing eigenvalues are 
\begin{equation}
M^2_{\pm}  =  \frac{1}{8} \left[ \left( \alpha v^2 + \beta v^2_\Phi \right) 
\pm \sqrt{ \left( \alpha v^2 + \beta v^2_\Phi \right)^2 - 4 v^2 v^2_\Phi \left ( \alpha \beta - \gamma^2 \right) }\right] \; ,
\end{equation}
where
\begin{eqnarray}
\alpha & =& g^2 + g^{\prime 2} + g_H^2 + 4 g_X^2 \; , \nn \\
\beta & = & g_H^2 + 4 g_X^2 \; , \nn \\
\gamma & = &  g_H^2 - 4 g_X^2  \; .
\label{eq:abg_def}
\end{eqnarray}

A strictly massless dark photon might not be phenomenologically desirable. 
One could have a Stueckelberg extension of the above model by including the Stueckelberg mass term 
\cite{Kors:2004dx,Kors:2005uz}
\begin{equation}
{\cal L}_{\rm Stu} = + \frac{1}{2} \left( \partial_\mu a + M_X X_\mu + M_Y B_\mu \right )^2 \; ,
\end{equation}
where $M_X$ and $M_Y$ are the Stueckelberg masses for the gauge fields $X_\mu$ and $B_\mu$ of 
$U(1)_X$ and $U(1)_Y$ respectively, and $a$ is the axion field.
Thus the neutral gauge boson mass matrix is modified as
\begin{equation}
\label{M1sq2}
{\mathcal M}_1^2 =     \begin{pmatrix}
\frac{g^{\prime 2} v^2 }{4} + M_Y^2 & - \frac{g^{\prime} g \, v^2 }{4}  &  \frac{g^{\prime} g_H v^2 }{4} & \frac{g^\prime g_X v^2}{2} + M_X M_Y \\
- \frac{g^{\prime} g \, v^2 }{4} & \frac{ g^2 v^2 }{4} & - \frac{g g_H v^2 }{4} 
& - \frac{ g g_X v^2  }{2} \\
 \frac{g^{\prime} g_H v^2 }{4} & - \frac{g g_H v^2 }{4}   & \frac{g^2_H  \lee v^2 + v^2_\Phi \rii }{4}  & 
 \frac{g_H g_X \lee v^2 - v^2_\Phi \rii }{2} \\ 
 \frac{g^\prime g_X v^2}{2} + M_X M_Y & - \frac{ g g_X v^2  }{2}
 & \frac{g_H g_X \lee v^2 - v^2_\Phi \rii }{2} & g_X^2 \left( v^2 + v^2_\Phi \right) + M_X^2
      \end{pmatrix} \; .
\end{equation}
It is easy to show that this mass matrix has only one zero mode corresponding to the photon, and three 
massive modes $Z,Z^\prime,Z^{\prime\prime}$. This mass matrix can be diagonalized by an orthogonal matrix.
The cubic equation for the three eigenvalues
can be written down analytically similar to solving the cubic equation for the vev $v_\De$ given in the Appendix~\ref{section:app}.
However their expressions are not illuminating and will not presented here.

As shown in Ref.~\cite{Kors:2005uz}, $M_Y$ will induce the mixing between $U(1)_Y$ and $U(1)_X$ and
the resulting massless eigenstate, the photon, will contain a $U(1)_X$ component, 
rendering the neutron charge, $Q_n= Q_u + 2 Q_d$, nonzero
unless $u$'s and $d$'s $U(1)_X$ charges are zero or proportional to their electric charges.
In this model, however, none of the two solutions can be satisfied.
Besides, left-handed SM fields are singlets under $U(1)_X$ while right-handed ones are charged.
It implies the left-handed and right-handed species may have different electric charges if the $U(1)_X$ charge 
plays a role on the electric charge definition.
Here we will set $M_Y$ to be zero to maintain the relations
$Q= I_3 + Y$ and $1/e^2= 1/g^{\prime 2} + 1/g^2$ same as the SM in order to avoid undesired features.
As a result, after making a rotation in the $1-2$ plane by the Weinberg angle $\th_w$, the mass matrix ${\mathcal M}_1^2$ can transform into a block diagonal matrix with the vanishing first column and first row. The nonzero 3-by-3 block matrix can be further diagonalized by an orthogonal matrix $\mathcal{O}$, characterized by three rotation angles $(\th_{12},\th_{23},\th_{13})$, 
\begin{align}
 \begin{pmatrix}
 Z_{SM} \\
 W^{\prime}_3 \\
 X
 \end{pmatrix}
 = \mathcal{O}(\th_{12},\th_{23},\th_{13}) \cdot
  \begin{pmatrix}
 Z \\
 Z^{\prime} \\
 Z^{\prime\prime}
 \end{pmatrix} \;\; \text{or} \;\;
   \begin{pmatrix}
 Z \\
 Z^{\prime} \\
 Z^{\prime\prime}
 \end{pmatrix}
 = \mathcal{O}^T(\th_{12},\th_{23},\th_{13}) \cdot
 \begin{pmatrix}
 Z_{SM} \\
 W^{\prime}_3 \\
 X
 \end{pmatrix} ,
\end{align}
where $Z_{SM}$ is the SM $Z$ boson without the presence of the $W^\prime_3$ and $X$ bosons.
In this model, the charged current mediated by the $W$ boson and the electric current by the photon $\ga$
are exactly the same as in the SM: 
\begin{align}
{\cal L}(\ga) &= \sum_f Q_f e \bar{f} \ga^\mu f A_\mu \;  , \nn \\
{\cal L}(W) &= \frac{g}{\sqrt{2}} \lee \overline{\nu_L} \ga^\mu e_L + \overline{u_L} \ga^\mu d_L \rii W^+_\mu + {\rm H.c.} \; ,
\end{align}
where $Q_f$ is the corresponding fermion electric charge in units of $e$.
On the other hand, neutral current interactions, including ones induced by $W^\prime$, take the following form~(for illustration,
only the lepton sector is shown but it is straightforward to include the quark sector)
\begin{equation}
{\cal L}_{NC} =  {\cal L}(Z) +{\cal L}(Z^\prime) + {\cal L}(Z^{\prime\prime}) +  {\cal L}(W^\prime) \; ,
\end{equation}
where 
\begin{align}
 {\cal L}(Z) & = g \mathcal{O}_{11} J^\mu_{\tiny{Z_{SM}}} + g_H \mathcal{O}_{21} J^\mu_{W^{\prime 3}} Z^{\prime}_\mu 
 +  g_X  \mathcal{O}_{31} J^\mu_X Z^{\prime\prime}_\mu \; ,  \nn \\
  {\cal L}(Z^\prime) & = g \mathcal{O}_{12} J^\mu_{\tiny{Z_{SM}}} + g_H \mathcal{O}_{22}  J^\mu_{W^{\prime 3}} Z^{\prime}_\mu 
 +  g_X \mathcal{O}_{32}  J^\mu_X Z^{\prime\prime}_\mu  \; , \nn \\ 
  {\cal L}(Z^{\prime\prime}) & = g \mathcal{O}_{13} J^\mu_{\tiny{Z_{SM}}} + g_H \mathcal{O}_{23} J^\mu_{W^{\prime 3}} Z^{\prime}_\mu 
 +  g_X \mathcal{O}_{33}  J^\mu_X Z^{\prime\prime}_\mu \; ,  \nn \\
 {\cal L}(W^\prime) &= \frac{g_H}{\sqrt{2}} \lee \overline{e^H_R} \ga^\mu e_R + \bar{\nu}_R \ga^\mu \nu^H_R  \rii W^{\prime p}_\mu + {\rm H.c.}  \; ,
\end{align}
and
\begin{align}
J^\mu_{\tiny{Z_{SM}}} & = \frac{1}{\cos\th_w}
\lee \lee \sum_{f= e,\nu} \overline{f_L} \ga^\mu (I_3 - Q_f \sin\th_w)  f_L + \overline{f_R} \ga^\mu (- Q_f \sin\th_w)  f_R \rii
+ \overline{e^H_R} \ga^\mu ( \sin\th_w) e^H_R \rii  \; , \nn\\
J^\mu_{W^{\prime 3}} & = \sum_{f=N_R, E_R}  \overline{f_R} \ga^\mu (I^H_3)  f_R \;  , \\
J^\mu_{X} & = \sum_{f=N_R, E_R} Q^f_X \overline{f_R} \ga^\mu f_R \; , \nn
\end{align}
with $I_3$ ($I^{H}_3$) being the $SU(2)_L$~($SU(2)_H$) isospin and $Q^f_X$ the $U(1)_X$ charge.
Detailed analysis of the implications of these extra gauge bosons is important 
and will be presented elsewhere.

%


\section{Phenomenology} \label{section:higgs_physics}

  
In this Section, we discuss some phenomenology implications of the model by examining 
the mass spectra of scalars and gauge bosons, $Z'$ constraints from various experiments, and Higgs properties of this model against
the LHC measurements on the partial decay widths of the SM Higgs boson, which is  $h_1$ in the model.


\subsection{Numerical Solutions for Scalar and Gauge Boson Masses}
\label{subsec:Nmh}

\begin{figure}[htbp!]
\centering
\includegraphics[clip,width=0.60\linewidth]{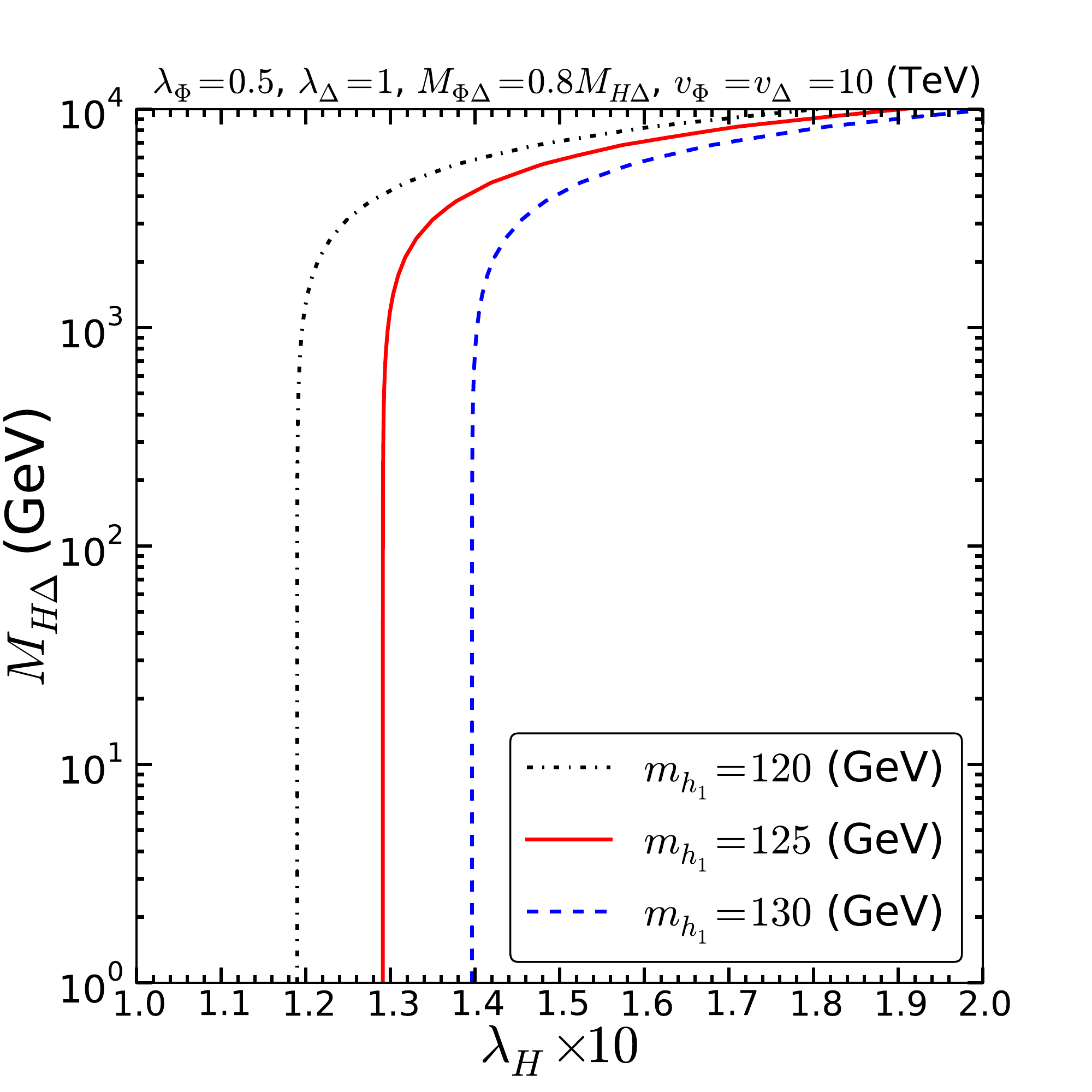}
\caption{ The SM Higgs mass dependence on $M_{H\De}$ and $\la_H$. 
In the limit of $v_\Phi$, $v_\De \gg v$, 
the Higgs mass is basically determined by two parameters $M_{H\De}$ and $\la_H$ only.
Other parameters are set as follows: $\la_\Phi=0.5$, $\la_{\Delta}=1$, $v_\Phi=v_\De=10\tev$ and 
$M_{\Phi\Delta}=0.8 M_{H\Delta}$. }
\label{fig:laH_mHD}  
\end{figure}

We first study the SM Higgs mass~($m_{h_1}$) dependence on the parameters in the mass matrix in Eq.~\eqref{eq:scalarbosonmassmatrix}.
As we shall see later the vev $v_\Phi$ has to be bigger than $10$ TeV~($\gg v =246$ GeV). 
In light of LEP measurements on the $e^+ e^- \to e^+ e^- $
cross-section~~\cite{LEP:2003aa}, the mass matrix will exhibit block-diagonal structure with the bottom-right 2-by-2 block much bigger than the rest and $h$ basically decouple from $\phi_2$ and $\delta_3$.

To demonstrate this behavior, we simplify the model by setting $\la_{H\De}$, $\la_{H\Phi}$, $\la_{\Phi\De}$ equal to zero
and then choose $\la_\Phi=0.5$, $\la_{\Delta}=1$, $v_\Phi=v_\De=10\tev$, and $M_{\Phi\Delta}=0.8 M_{H\Delta}$ so that
one can investigate how the Higgs mass varies as a function of $\la_H$ and $M_{H\De}$.
As shown in Fig.~\ref{fig:laH_mHD}, when $M_{H\De}$ is very small compared to $v_\Phi$, the Higgs mass is simply the (1,1)
element of the mass matrix, $2 \la_H v^2$, and $h_1$ is just $h$, {\it i.e.}, $O_{11}^2 \simeq 1$. Nonetheless, when $M_{H\De}$ becomes comparable to $v_\Phi$, and the (1,2) element of the mass matrix gives rise to a sizable but negative contribution to the Higgs mass, requiring  a larger value of $\la_H$ than the SM one so as to have a correct Higgs mass. In this regime, $\vert O_{11} \vert$ is, however, still very SM-like: $O_{11}^2 \simeq 1$. Therefore, one has to measure the quartic coupling $\la_H$ 
through the double Higgs production to be able to differentiate this model from the SM.          

For the analysis above, we neglect the fact all vevs, $v_\Phi$, $v_\De$ and $v$ are actually functions of the parameters 
$\mu$s, $M$s and $\la$s in Eq.~\eqref{eq:higgs_pot}, the total scalar potential. The analytical solutions of the vevs are collected in Appendix~\ref{section:app}.
As a consequence, we now numerically diagonalized the matrices~\eqref{eq:scalarbosonmassmatrix}
and \eqref{goldstonemassmatrix}
as functions of $\mu$, $M$ and $\la$, {\it i.e.}, replacing all vevs by the input parameters.
It is worthwhile to mention that $v_\Delta$ has three solutions as Eq.~\eqref{vevdelta} is a cubic equation for $v_\Delta$.
Only one of the solutions corresponds to the global minimum of the full scalar potential. We, however, include both the global
and local minimum neglecting the stability issue for the latter since it will demand detailed study on the nontrivial potential shape
 which is beyond the scope of this work. 

\begin{figure}[htbp!]
\centering
\includegraphics[clip,width=0.6\linewidth]{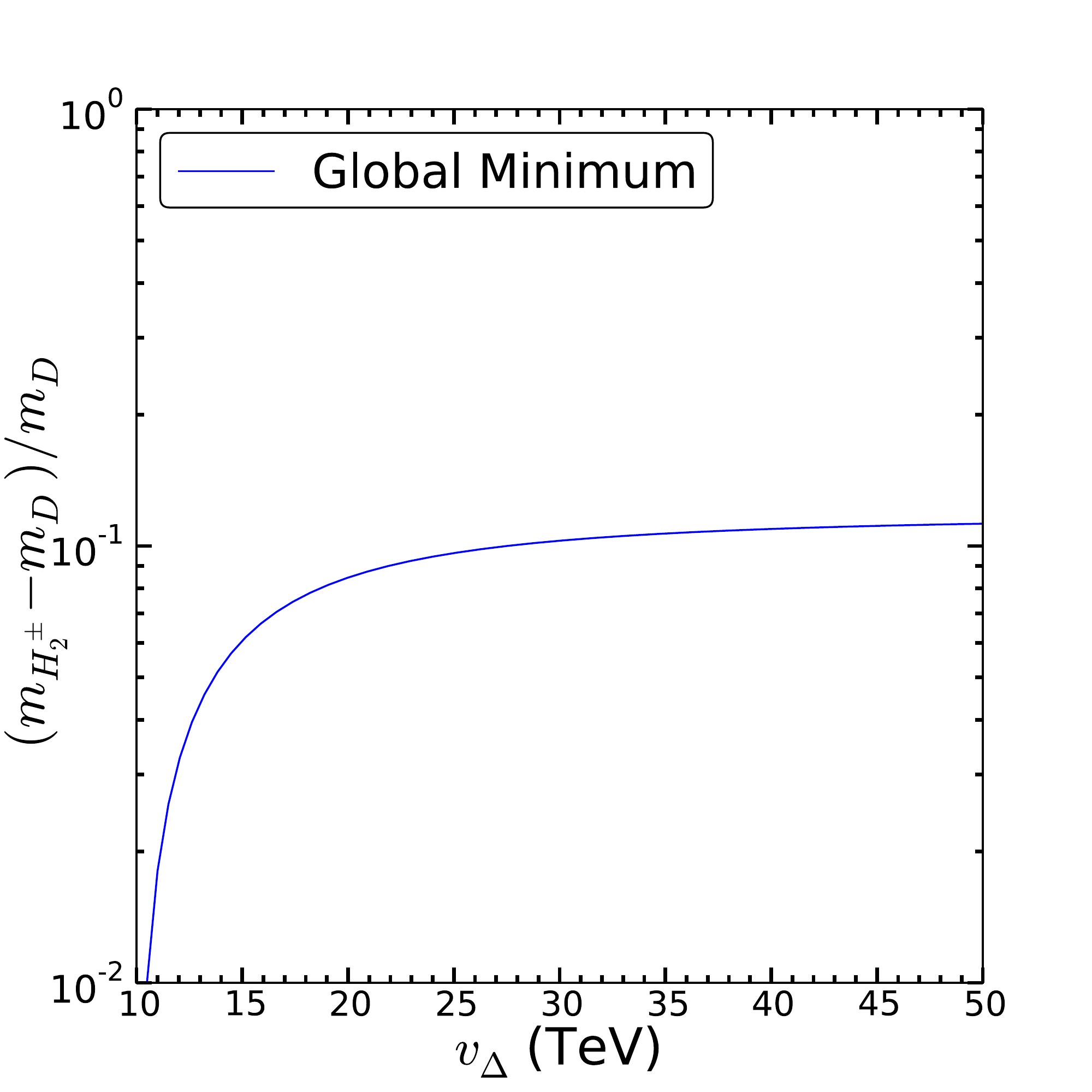}
\caption{Mass difference of the charged Higgs and dark matter normalized to $m_{D}$ 
as a function of $v_\De$ for $M_{\Phi\Delta}=0.8 M_{H\Delta}$ and 
$v_\Phi= 10$ TeV used in Table~\ref{tab:Bench_points}.
}\label{fig:spliting}  
\end{figure}
\begin{table}[htp!]
\begin{tabular}{|c||c|c|c||c|c|c|}
\hline
\hline
      \multicolumn{1}{|c||}{} &
      \multicolumn{3}{c||}{\texttt{Global}} &
      \multicolumn{3}{c|}{\texttt{Local}} \\
\hline
Benchmark points &  A & B & C  & D & E & F \\
\hline
$\la_{H\De}$ & $0.186$ &$1.390$  & $0.186$ & $2.406$ & $0.557$ &$2.406$ \\
\hline
$\la_{H\Phi}$  & $0.669$ &$1.390$  & 7.467$\times 10^{-2}$ & $2.004$ & 5.181$\times 10^{-2}$ &$0.557$  \\
\hline
$\la_{\Phi\De}$ & $1.390$ &$1.390$  & 3.594$\times 10^{-2}$ & 2.077$\times 10^{-2}$ & $1.390$ &$0.386$  \\
\hline
$\frac{v_{\Delta}}{\tev}$ & $10.993$ &$10.993$  & $12.050$ & $10.5$ & $15.870$ &$10.993$  \\
\hline
$\la_H$ & $4.456$ &$3.155$  & $0.141$ & $3.540$ & $1.775$ &$1.582$  \\
\hline
$\frac{M_{H\Delta}}{\gev}$ &$24.632$ &$59.021$  & $1.0$ & $1.0$ & $78.979$ &$18.408$  \\ 
\hline
\hline
$\frac{m_{h1}}{\gev}$ &$125.133$ &$124.988$  & $125.384$ & $125.049$ & $124.953$ &$125.073$  \\ 
$\frac{m_{h2}}{\tev}$ &$1.717$ &$1.692$  & $9.995$ & $10.009$ & $1.783$ &$9.394$  \\ 
$\frac{m_{h3}}{\tev}$ &$1.842$ &$18.420$  & $17.044$ & $14.857$ & $24.513$ &$15.926$  \\ 
$\frac{m_{D}}{\gev}$ &$511.162$ &$791.244$  & $106.287$ & $101.447$ & $1049.883$ &$441.884$  \\ 
$\frac{m_{H^\pm}}{\gev}$ &$520.375$ &$805.504$  & $109.773$ & $102.469$ & $1119.586$ &$449.848$  \\ 
$\mathcal{O}_{11}^2$ &$0.823$ &$0.897$  & $0.999$ & $0.996$ & $0.939$ &$0.999$  \\ 
$R_{\gamma\gamma}$ &$0.820$ &$0.895$  & $0.920$ & $0.903$ & $0.938$ &$0.995$  \\ 
$R_{\gamma Z}$ &$0.822$ &$0.896$  & $0.968$ & $0.958$ & $0.938$ &$0.997$  \\
\hline
\end{tabular}
\caption{Six representative benchmark points.  
We fix other parameters as $\la_\Phi=0.5$, $\la_{\Delta}=1$, $v_\Phi=10\tev$, and 
$M_{\Phi\Delta}=0.8 M_{H\Delta}$.
}
\label{tab:Bench_points}
\end{table}

In order to explore the possibility of a non-SM like Higgs having a 125 GeV mass, we further allow for nonzero mixing couplings.
We perform a grid scan with 35 steps of each dimension in the range 
\begin{eqnarray}
\label{domain}
10^{-2} &\leq \, \la_{H\De}   \, \leq &  5 \; , \nonumber\\
10^{-2} &\leq \, \la_{H\Phi} \, \leq &   5 \; , \nonumber\\
10^{-2} &\leq \, \la_{\Phi\De} \, \leq & 5 \; , \nonumber\\
10^{-1} &\leq \, \la_{H} \, \leq & 5 \; , \nonumber\\
1.0 &\leq \, M_{H\Delta}/\gev \, \leq & 2\times 10^4 \; , \nonumber\\
1.05 &\leq \, v_{\Delta}/v_{\Phi} \, \leq & 5.0 \; .
\end{eqnarray} 
In order not to overcomplicate the analysis, from now on we make $\la_\Phi=0.5$, $\la_{\Delta}=1$, 
$v_\Phi=10\tev$, and $M_{\Phi\Delta}=0.8 M_{H\Delta}$, unless otherwise stated. 
In Table~\ref{tab:Bench_points}, we show 6 representative benchmark points~(3 global and 3 local minima)
from our grid scan with the dark matter mass $m_D$ and the charged Higgs $m_{H^\pm}$ of order few hundred GeVs,
testable in the near future.
It is clear that \texttt{Global} scenario can have the SM Higgs 
composition significantly different from 1, as $\mathcal{O}_{11}^2\sim 0.8$ in benchmark point A, but $h_1$ is often just $h$ in \texttt{Local} case. On the other hand, the other two heavy Higgses are as heavy as TeV because their mass are basically determined
by $v_\De$ and $v_\Phi$. 
  
For the other mass matrix Eq.~\eqref{goldstonemassmatrix}, we focus on the mass splitting between the dark matter and the charged Higgs and it turns out the mass splitting mostly depends on $v_\Phi$ and $v_\De$.   
In Fig.~\ref{fig:spliting}, 
we present the mass difference normalized to $m_{D}$ 
as a function of $v_\De$ for $M_{\Phi\Delta}=0.8 M_{H\Delta}$ and 
$v_\Phi= 10$ TeV used in Table~\ref{tab:Bench_points}. 
The behavior can be easily understood as
\begin{align}
\frac{m_{H^\pm_2} - m_D }{ m_{D} } & \simeq \frac{ v^2_\De - v^2_\Phi + \vert v^2_\De - v^2_\Phi \vert  }{ 8 \, v^2_\De + 2 \, v^2_\Phi } \nn\\
&=   \frac{1 -  v^2_\Phi/v^2_\De}{4 +  v^2_\Phi/v^2_\De}
\end{align}
where we have neglected terms involving $v$ since we are interested in the limit of $v_\De$, $v_\Phi \gg v$.
Note that this result is true as long as $M_{\Phi\Delta}=0.8 M_{H\Delta}$ and $v_\De > v_\Phi \gg v$
regardless of the other parameters.


\subsection{$Z'$ Constraints}

\begin{figure}[htbp!]
\centering
\includegraphics[clip,width=0.5\linewidth]{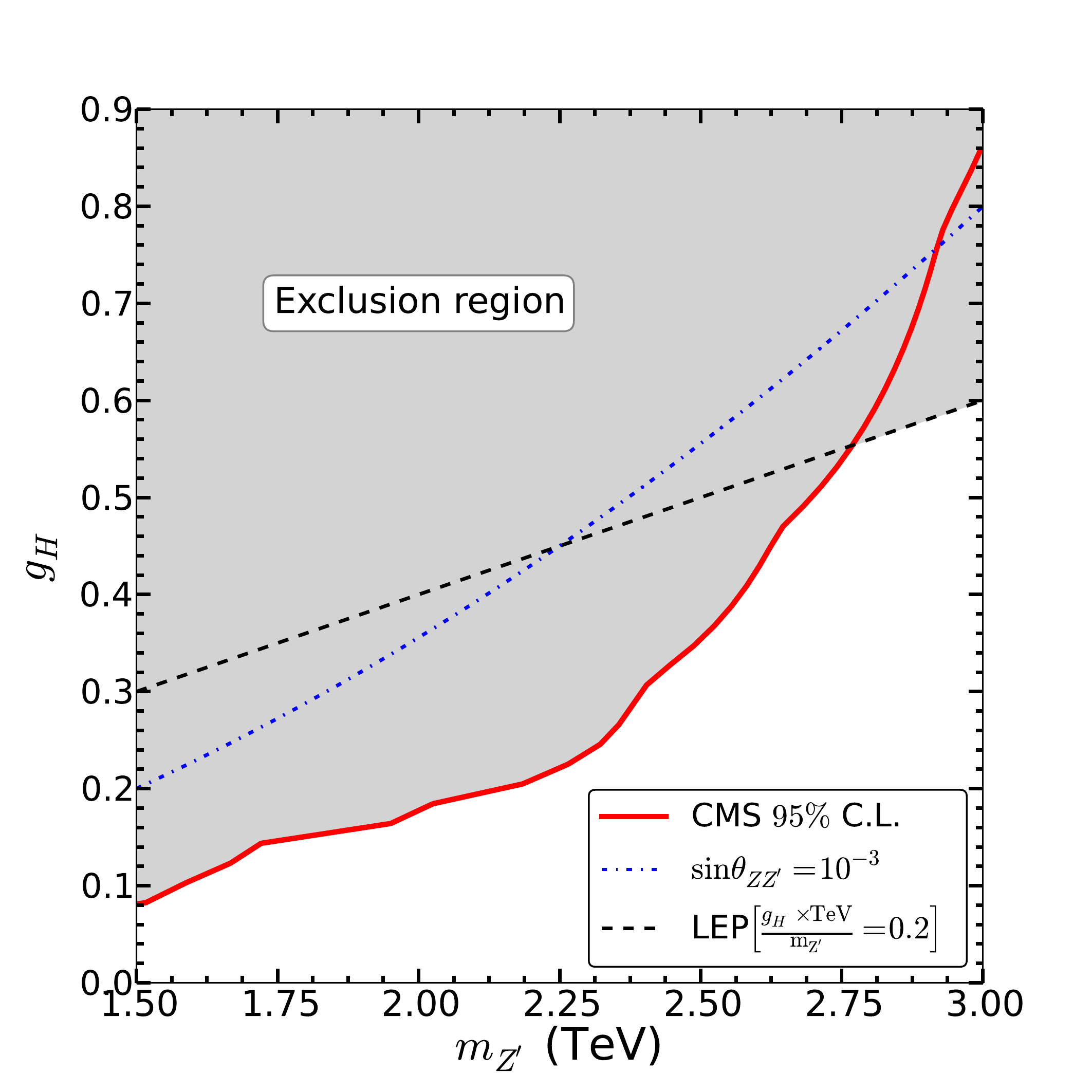}
\caption{Constraints from the direct $Z^\prime$ resonance search based on dilepton channels~(red line) at the LHC,
from $Z-Z^\prime$ mixing~(blue dotted line) and the LEP constraints on the electron-positron scattering cross-section~(black dashed line) in the $m_{Z^\prime}-g_H$ plane.
}
\label{fig:gH_mzp}  
\end{figure}

Since performing a full analysis of the constraints on extra neutral gauge bosons including all possible 
mixing effects from $\gamma, Z,Z',Z''$ is necessarily complicated, we will be contented in this work 
by a simple scenario discussed below.

The neutral gauge boson mass matrix in Eq.~(\ref{M1sq1}) can be simplified a lot by making $U(1)_X$ a global symmetry.
In the limit of $g_X=M_X=0$, the $U(1)_X$ gauge boson $X$ decouples from the theory, and the 3-by-3 mass matrix 
in the basis of $B$, $W^3$ and $W^{\prime 3}$ can be rotated into the mass basis of the SM $\ga$, $Z$ and the $SU(2)_H$ $Z^\prime$ by:
\begin{align}
      \begin{pmatrix}
m^2_\ga & 0 & 0  \\
0 & m^2_Z & 0  \\
0 & 0 & m^2_{Z^\prime}  \\ 
      \end{pmatrix} =  
R_{23} \lee \th_{ZZ^\prime} \rii^T
R_{12}\lee \th_w \rii^T
     \begin{pmatrix}
\frac{g^{\prime 2} v^2 }{4} & - \frac{g^{\prime} g \, v^2 }{4}  &  \frac{g^{\prime} g_H v^2 }{4}  \\
- \frac{g^{\prime} g \, v^2 }{4} & \frac{g^2 v^2 }{4} & - \frac{g g_H v^2 }{4}  \\
 \frac{g^{\prime} g_H v^2 }{4} & - \frac{g g_H v^2 }{4} & \frac{g^2_H \lee v^2 + v^2_\Phi \rii}{4}  \\ 
      \end{pmatrix}
R_{12}\lee \th_w \rii
R_{23} \lee \th_{Z Z^\prime} \rii  ,
\end{align}
where $R_{ij}$ refers to a rotation matrix in the $i-j$ block; {\it e.g.}, $R_{12} \lee \th_w \rii$ is a rotation along the $z$-axis ($W^{\prime 3}$)
direction with $\cos\th_w$ as the $(1,1)$ and $(2,2)$ element and $-\sin\th_w$ as the $(1,2)$ element ~($\sin\th_w$ for $(2,1)$).
The mixing angles can be easily obtained as
\begin{align}
\sin\th_w &= \frac{ g^\prime }{ \sqrt{g^2 + g^{\prime 2}}  }    \;\; ,  \;\; \\
\sin\th_{Z Z^\prime}&=  \frac{ \sqrt{2} \sqrt{g^2 + g^{\prime 2} } \, g_H  \, v^2  } 
{ \kappa^{1/4} \lee - (g^2 + g^{\prime 2})v^2 + g_H^2 \lee v^2+ v^2_\Phi \rii + \kappa^{1/2} \rii^{1/2} } \;\; , 
\label{eq:ZZp_mixing}
\end{align}
where $\kappa= \left( \left( g^2 + g^{\prime 2} + g_H^2 \right)  v^2 + g_H^2 v^2_\Phi \right)^2 
- 4 g_H^2 \left( g^2 + g^{\prime 2} \right) v^2 v^2_ \Phi $.
In the limit of $v_\Phi \gg v$, we have the approximate result
\begin{equation}
\sin\th_{Z Z^\prime} \approx  \frac{ \sqrt{g^2 + g^{\prime 2} }  \, v^2  } { g_H \, v^2_\Phi }  \; ,
\end{equation}
and
 \begin{align}
 m_Z \approx \sqrt{g^2 + g^{\prime 2} } \frac{v}{2} \;\; , \;\; m_{Z^\prime} \approx g_H \frac{v_\Phi}{2} .
 \end{align}
A couple of comments are in order here. 
First, the SM Weinberg angle characterized by $\th_w$ is unchanged in the presence of $SU(2)_H$.
Second, the vev ratio $v^2/v_\Phi^2$ controls the mixing between the SM $Z$ and $SU(2)_H$ $Z^\prime$.
However, the $Z-Z^\prime$ mixing for TeV $Z^\prime$ is constrained to be roughly less than $0.1\%$, results from $Z$ resonance
line shape measurements~\cite{ALEPH:2005ab}, electroweak precision test (EWPT) data~\cite{Erler:2009jh} and
collider searches via the $W^+W^-$ final states~\cite{Andreev:2012zza,Andreev:2014fwa}, depending on underlying models.

Direct $Z^\prime$ searches based on dilepton channels at the LHC~\cite{Chatrchyan:2012oaa,Aad:2014cka,Khachatryan:2014fba} yield stringent constraints on the mass of $Z^\prime$ of this model since right-handed SM fermions which are part of $SU(2)_H$
doublets couple to the $Z^\prime$ boson and thus $Z^\prime$ can be produced and decayed into dilepton at the LHC.
To implement LHC $Z^\prime$ bounds, we take the $Z^\prime$ constraint~\cite{Khachatryan:2014fba} on the Sequential Standard Model~(SSM)
with SM-like couplings~\cite{Altarelli:1989ff}, rescaling  by a factor of $g^2_H(\cos^2 \th_w/g^2 ). $ 
It is because first $SU(2)_H$ 
does not have the Weinberg angle in the limit of $g_X=0$ and second
we assume $Z^\prime$ decays into the SM fermions only and branching ratios into the heavy fermions
are kinematically suppressed, {\it i.e.}, $Z^\prime$ decay branching ratios are similar to those of the SM $Z$ boson.
The direct search bound becomes weaker once heavy fermion final states are open.  
Note also that $Z^\prime$ couples only to the right-handed SM fields unlike $Z^\prime$ in the SSM couples 
to both left-handed and right-handed fields as in the SM.
 The SSM $Z^\prime$ left-handed couplings are, however,
dominant since the right-handed ones are suppressed by the Weinberg angle.  Hence we simply rescale the SSM result
by $g^2_H (\cos^2 \th_w/g^2 )$ without taking into account the minor difference on the chiral structure of the couplings. 

In addition, $Z^\prime$ also interacts with the right-handed electron and will contribute to $e^+ e^- \to \ell^+ \ell^-$ processes.
LEP measurements on the cross-section of $e^+ e^- \to \ell^+ \ell^-$ can be translated into the constraints on the new physics scale in the context of the effective four-fermion interactions~\cite{LEP:2003aa}
\begin{align}
\mathcal{L}_{\text{eff}}= \frac{4\pi}{ \lee1 + \de \rii \Lambda^2} \sum_{i,j=L,R} \eta_{i,j} \bar{e}_i \ga_\mu e_i \bar{f}_j \ga^\mu f_j \, ,
\end{align}
where $\de=0~(1)$ for $f\neq e~(f=e)$ and $\eta_{ij}=1~(-1)$ corresponds to constructive~(destructive) interference between the SM and the new physics processes. 
On the other hand, in our model for $m_{Z^\prime} \sim $TeV the contact interactions read
\begin{align}
\mathcal{L}_{\text{eff}}= - \lee 1 + \de\rii \frac{g^2_H}{ m^2_{Z^\prime}}  \bar{e}_R \ga_\mu e_R \bar{f}_R \ga^\mu f_R \, .
\end{align}
It turns out the strongest constraint arises from $e_L^+ e_R^- \to e_L^+ e_R^-$ with $\La=8.9$ TeV and $\eta=-1$~\cite{LEP:2003aa}, which implies
\begin{align}
\frac{g_H}{m_{Z^\prime} } \lesssim \frac{0.2}{\text{TeV}} \, \;\; \text{and} \;\;
v_\Phi  \gtrsim 10 \, \text{TeV} \, .
\end{align}

In Fig.~\ref{fig:gH_mzp}, in the plane of $g_H$ and $m_{Z^\prime}$, we show the three constraints: the $Z-Z^\prime$ mixing by the blue dotted line, the heavy narrow dilepton resonance searches from CMS~\cite{Khachatryan:2014fba} in red and the LEP bounds on the electron-positron scattering cross-section of $e^+ e^- \to e^+ e^-$~\cite{LEP:2003aa} in black, 
where the region above each of the lines is excluded.
The direct searches are dominant for most of the parameter space of interest, while the LEP constraint becomes most stringent toward the high-mass region. From this figure,
we infer that for $0.1 \lesssim g_H \lesssim 1$, the $Z^\prime$ mass has to be of order $\mathcal{O}$(TeV). 
Note that $m_{Z^\prime} \approx g_H v_\Phi/2$ and 
it implies $v_\Phi \sim 30 $ TeV for $m_{Z^\prime} \sim 1.5$ TeV but $v_\Phi$ can be 10  TeV
for $m_{Z^\prime} \gtrsim 2.75$ TeV.    

 \subsection{Constraints from the 125 GeV SM-like Higgs}

\begin{figure}[htp!]
\centering
\includegraphics[clip,width=0.46\linewidth]{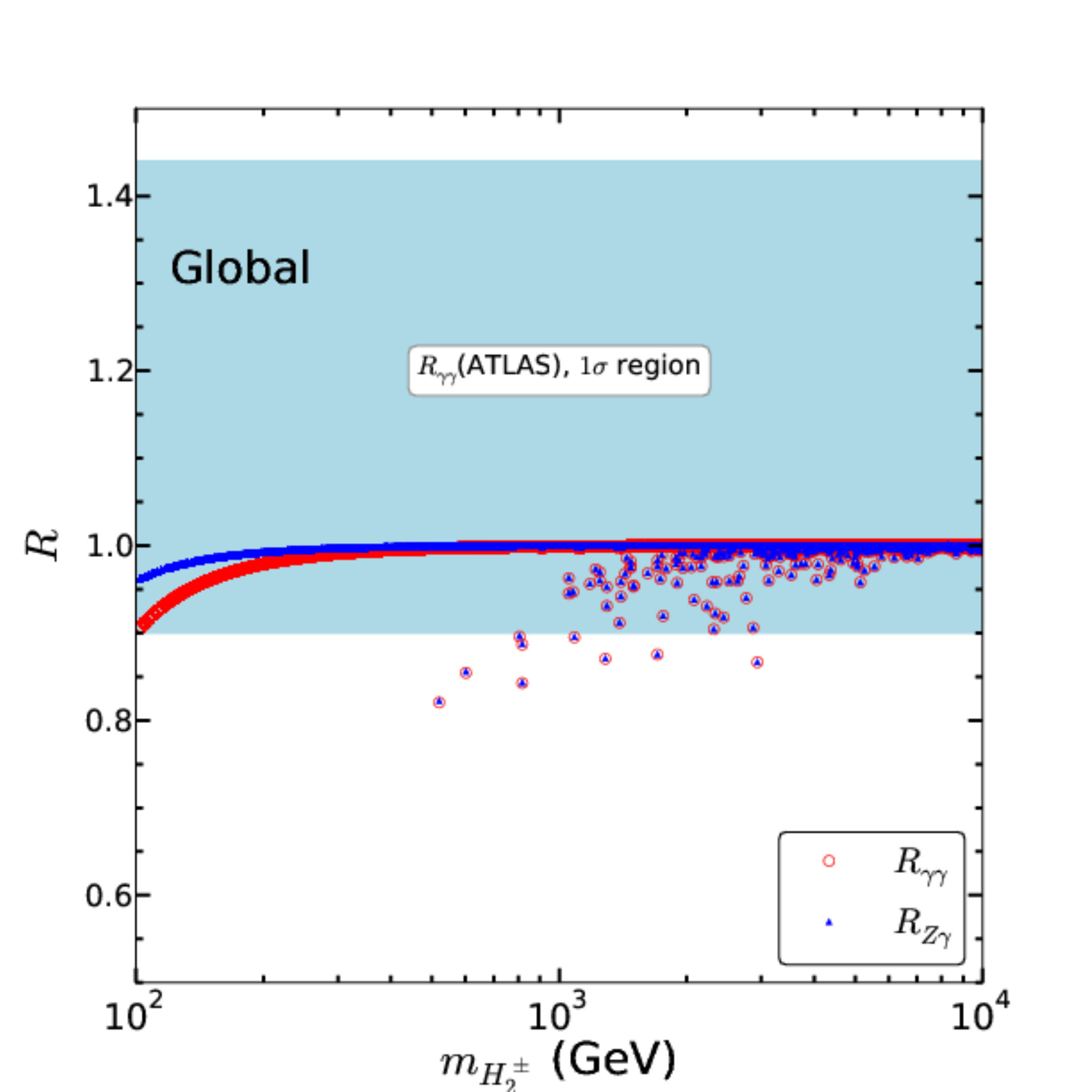}
\includegraphics[clip,width=0.46\linewidth]{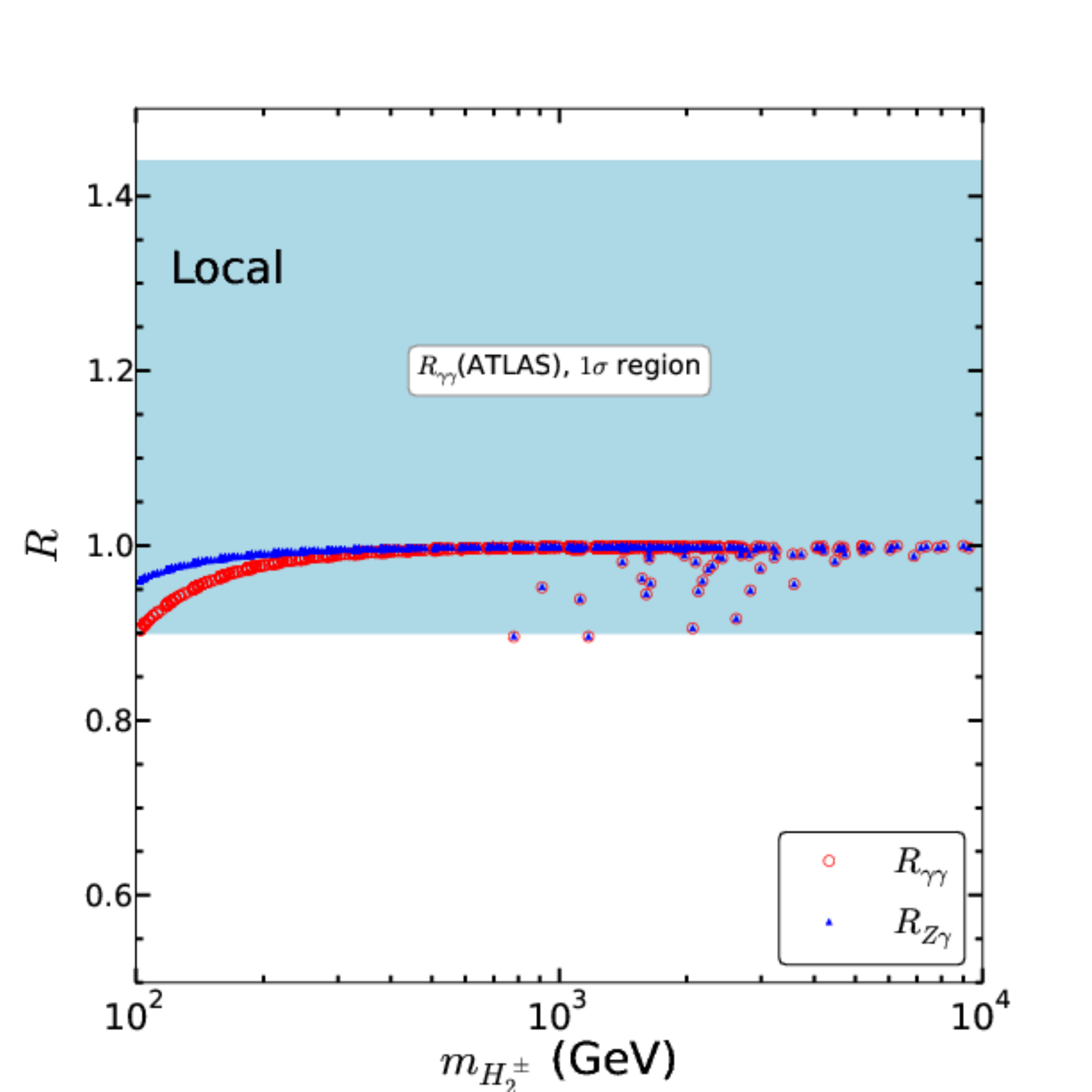}
\caption{Predictions of $h \to \ga\ga$ and $h \to Z \ga$ in this model. Due to the fact the $\vert O_{11} \vert \sim 1$ and $\la_H$ is always positive,
$R_{\ga\ga}$ is often less than the SM prediction while $R_{Z\ga}$ ranges from $0.9$ to $1$, given the ATLAS and CMS measurements on $R_{\ga\ga}$: 
: $1.17 \pm 0.27$ (ATLAS ~\cite{Aad:2014eha}) and
 $1.13 \pm 0.24$ (CMS ~\cite{CMS:2014ega}).
Only ATLAS result is shown which completely covers the CMS $1\sig$ confidence region.  
The left~(right) panel is for \texttt{Global}~(\texttt{Local}) scenario.
}
\label{fig:hZga}  
\end{figure}

 We begin with the SM Higgs boson $h_1$ partial decay widths into SM fermions. Based on the Yukawa couplings in Eqs.~\eqref{eq:Yuk_Q},
\eqref{eq:Yuk_Q1} and \eqref{eq:Yuk_L}, only the flavor eigenstate $H_1$ couples to two SM fermions. Thus, the coupling of the mass eigenstate $h_1$
to the SM fermions is rescaled by $O_{11}$, making the decay widths universally reduced by $ O_{11}^2 $ compared to the SM, 
which is different from generic 2HDMs.

For the tree-level (off-shell) $h_1 \to W^+ W^-$, since the SM $W$ bosons do not mix with additional $SU(2)_H$~(and $U(1)_X$) gauge bosons and $\phi_2$ and $\delta_3$ are singlets under the $SU(2)_L$ gauge group,
the partial decay width is also suppressed  
by $ O_{11}^2 $.
On the other hand, $h_1 \to Z Z$ receives additional contributions from the $Z-Z'$ mixing and thus the $\delta_3$ and $\phi_2$ components charged under $SU(2)_H$ will also contribute.
The mixing, however, are constrained to be small~($ \lesssim 10^{-3}$) 
by the EWPT data and LEP measurement on the electron-positron scattering cross-section as discussed in previous subsection.
We will not consider the mixing effect here and 
the decay width is simply reduced by $ O_{11}^2 $ identical to other tree-level decay channels.

Now, we are in a position to explore the Higgs radiative decay rates into two photons and one $Z$ boson and one photon,
normalized to the SM predictions. 
For convenience, we define $R_{XX}$ to be the production cross section of an SM Higgs boson decaying to $XX$
 divided by the SM expectation as follows:
\begin{equation}\label{eq:ratioR}
R_{XX}\equiv \frac{\sigma(pp\to h_1)}{\sigma_{\rm SM}(pp\to h_1)}
\frac{{\rm Br}(h_1\to XX)}{{\rm Br}_{\rm SM}(h_1\to XX)} \; .
\end{equation}
Note that first additional heavy colored fermions do not modify the Higgs boson production cross section, especially via $gg \to h_1$,
because the $SU(2)_H$ symmetry forbids the coupling of $ h_1 \bar{f} f$, where $f$ is the heavy fermion. 
Second, in order to avoid the observed Higgs invisible decay constraints in the vector boson fusion production mode: Br($h_1 \to \text{ invisible})<0.29$~\cite{ATLAS:2015inv,CMS:2015dia}, we constrain ourself to the region of 
$2m_{D} > m_h$ so that the Higgs decay channels are the same as in the SM. 
As a consequence, Eq.~\eqref{eq:ratioR} becomes,
\begin{equation}\label{eq:Ratio2}
R_{\gamma\gamma}=\frac{\Gamma(h_1\to \gamma\gamma)}{\Gamma_{\rm SM}(h_1\to \gamma\gamma)} \;\; , \;\;
R_{Z \gamma}=\frac{\Gamma(h_1\to \gamma Z)}{\Gamma_{\rm SM}(h_1\to \gamma\gamma)} \;\; ,
\end{equation}
similar to the situation of IHDM.

As mentioned before due to the existence of $SU(2)_H$, there are no terms involving one SM Higgs boson and two heavy
fermions, which are not $SU(2)_H$ invariant. Therefore, the only new contributions to $h_1 \to \ga\ga$ and $h_1 \to Z\ga$ arise
from the heavy charged Higgs boson, $H^{\pm}_2$. In addition, $h_1$ consists of $\delta_3$ and $\phi_2$ apart from $H^0_1$.
With the quartic interactions 
$2\la_H H^\dag_2 H_2 H^\dag_1 H_1 + \la_{H\De} H^\dag_2 H_2 \De^2_3/2 
+ \la_{H\Phi} H^\dag_2 H_2 \Phi^*_2 \Phi_2$, there are in total three contributions
to the $H^{\pm}_2$-loop diagram.    
The Higgs decay widths into $\ga\ga$ and $Z\ga$ including new scalars
can be found in Refs.~\cite{Gunion:1989we,Djouadi:2005gi,Djouadi:2005gj,Chen:2013vi}
and they are collected in Appendix~\ref{section:app-HDecays} for convenience.

The results of $h \to \ga\ga$~(red circle) and $h \to Z \ga$~(blue square) are presented 
in Fig.~\ref{fig:hZga} for  both the 
\texttt{Global} minimum case (left) and additional points which 
having the correct Higgs mass only at \texttt{Local} minimum (right).
All the scatter points were selected from our grid scan described in subsection~\ref{subsec:Nmh}.
It clearly shows that the mass of the heavy charged Higgs has to be larger than $100$ GeV in order to satisfy
the LHC measurements on $h \to \ga\ga$~\cite{Aad:2014eha,CMS:2014ega} while the corresponding $h \to Z \ga$ ranges from
$0.9$ to $1$.
Unlike normal IHDM where $\la_3$ in $\la_3  |H_1|^2 \,  |H_2|^2 $ can be either positive or negative, in this model
we have $\la_H \lee |H_1 |^2 + |H_2|^2 \rii^2$, where $\la_H$ as a quartic coupling has to be positive to ensure the potential is bounded from below.
It implies that for $\vert O_{11}\vert^2$ being very close to 1 like the benchmark point C in Table~\ref{tab:Bench_points}, $h_1$ is mostly $h$, the neutral component in $H_1$, and $R_{\ga\ga}$ is determined by $\la_H$.
In this case, $R_{\ga\ga}$, corresponding to the red curve, is always smaller than 1 in the region of interest where 
$ m_{H^\pm_2} > 80$ GeV ~(due to the LEP constraints on the charged Higgs in the context of IHDM~\cite{Pierce:2007ut}), 
while $R_{Z\ga}$ denoted by the blue curve also has values below the SM prediction as well.

In contrast, there are some points such as the benchmark point A where the contributions from $\phi_2$ and $\delta_3$ are significant,
they manifest in Fig.~\ref{fig:hZga} as scatter points away from the lines but also feature $R_{\ga\ga}$, $R_{Z\ga} < 1$.
 

\subsection{Dark Matter Stability}\label{section:DM_decay}

In G2HDM, although there is no residual (discrete) symmetry left from $SU(2)_H$ symmetry breaking, the lightest particle among the heavy $SU(2)_H$ fermions~($u^H$, $d^H$, $e^H$, $\nu^H$), the $SU(2)_H$ gauge boson
$W^\prime$~(but not $Z^\prime$ because of $Z-Z^\prime$ mixing)
and the second heaviest eigenstate in the mass matrix of Eq.~\eqref{goldstonemassmatrix}\footnote{The lightest one being the Goldstone boson absorbed by $W^\prime$.} is stable  
due to the gauge symmetry and the Lorentz invariance for the given particle content.
In this work, we focus on the case of the second heaviest eigenstate in the mass matrix of Eq.~\eqref{goldstonemassmatrix},
a linear combination of $H_2^{0*}$, $\De_p$ and $G^p_H$, being  DM by assuming the $SU(2)_H$ fermions and $W^\prime$ are heavier than it.

At tree level, it is clear that all renormalizable interactions always have even powers of the potential DM candidates in the set $D~\equiv \{ u^H, d^H, e^H, \nu^H, W^\prime  ,H_2^{0*}, \De_p , G^p_H\}$.
It implies they always appear in pairs and will not decay solely into SM particles. 
In other words, the decay of a particle in $D$ must be accompanied by the production of another particle in $D$, rendering the lightest
one among $D$ stable.

Beyond the renormalizable level, one may worry that radiative corrections involving DM will create non-renormalizable interactions,
which can be portrayed by high order effective operators, and lead to the DM decay.
Assume the DM particle can decay into solely SM particles via certain higher dimensional operators, which conserve the gauge symmetry
but spontaneously broken by the vevs of $H_1$, $\Phi_H$, $\De_H$.
The SM particles refer to SM fermions, Higgs boson $h$ and $\phi_2$~(which mixes with $h$)\footnote{The SM gauge bosons are not included since they decay into SM fermions, while inclusion of $\de_3$ will not change the conclusion because it carries zero charge in term of the quantum numbers in consideration.}. The operators, involving one DM particle and the decay products,
are required to conserve the gauge symmetry. We here focus on 4 quantum numbers: $SU(2)_L$ and $SU(2)_H$ isospin,
together with $U(1)_Y$ and $U(1)_X$ charge. DM is a linear combination of three flavour states: $H_2^{0*}$, $\De_p$ and $G^p_H$.
First, we study operators with $H_2^{0*}$ decaying leptonically into $n_1$ $\nu_L$s, $n_2$ $e_L$s, $n_3$ $\nu_R$s,
$n_4$ $e_R$s, $n_5$ $h$s and $n_6$ $\phi_2$s\,\footnote{The conclusion remains the same if quarks are also included.}. 
One has, in terms of the quantum numbers $(I^H_3,I_3,Y,X)$,
\begin{align}
(1/2, 1/2, - 1/2, -1) &= n_1*(0, 1/2, -1/2, 0) + n_2* ( 0, -1/2, -1/2, 0)  \nonumber \\
& +  n_3*(1/2, 0, 0, 1)+ n_4*(-1/2, 0, -1, -1) \\
& +  n_5* (1/2, -(1/2), 1/2, 1) + n_6* ( -1/2, 0, 0, 1 ) \; ,  \nonumber
\end{align}
 which gives
 \begin{align}
n_3= 1-n_1 \;\; , \;\; n_4= - n_2 \;\; ,\;\; n_5= -1 +n_1 -n_2 \;\; , \;\; n_6 = -1 \; \; .  
\end{align}
This implies the number of net fermions (fermions minus anti-fermions) is $n_1 + n_2+ n_3 + n_4=1$,
where positive and negative $n_i$ correspond to fermion and anti-fermion respectively. 
In other words, if the number of fermions is odd, the number of 
anti-fermions must be even; and vice versa.
Clearly, this implies Lorentz violation since the total fermion number (fermions plus anti-fermions)
is also odd.
Therefore, the flavor state $H^{0*}_2$ can not decay solely into SM particles.
It is straightforward to show the conclusion applies to $\De_p$ and $G^p_H$ as well.
As a result, the DM candidate, {\it a superposition of $H_2^{0*}$, $\De_p$ and $G^p_H$},
is stable as long as it is the lightest one among the potential dark matter candidates $D$.

Before moving to the DM relic density computation, we would like to comment on effective operators with an odd number
of $D$, like three $D$s or more. It is not possible that this type of operators, invariant under the gauge symmetry, will induce
operators involving only one DM particle, by connecting an even number of $D$s into loop. 
 It is because those operators linear in $D$ violate either the gauge symmetry or Lorentz invariance and hence can never be (radiatively) generated from operators obeying the symmetries. After all, the procedure of reduction on the power of $D$, {\it i.e.}, closing loops with proper vev insertions (which amounts to adding Yukawa terms), also complies with the gauge and Lorentz symmetry. 

\subsection{Dark Matter Relic Density}\label{section:DM_relic}

We now show this model can reproduce the correct DM relic density.
As mentioned above, the DM particle will be a linear combination of $G^p_H$, $\De_p$ and $H_2^{0*}$. 
Thus (co)-annihilation channels are relevant if their masses are nearly degenerated and the analysis can be quite involved. 
In this work, we constrain ourselves in the simple limit where the G2HDM becomes IHDM, in which case 
the computation of DM relic density has been implemented
in the software package {{\tt micrOMEGAs}}~\cite{Belanger:2008sj,Belanger:2010pz}. For a recent detailed analysis of IHDM,
see Ref.~\cite{Arhrib:2013ela} and references therein. 
In other words, we are working in the scenario that the DM being mostly $H_2^0$ and 
the SM Higgs boson $h$ has a minute mixing with $\delta_3$ and $\phi_2$.  
The IHDM Higgs potential reads,
\begin{eqnarray}
\label{IHDM_pot}
V_{\rm IHDM} &=& \mu_1^2 |H_1|^2 + \mu_2^2 |H_2|^2 + \lambda_1 |H_1|^4
+ \lambda_2 |H_2|^4 +  \lambda_3 |H_1|^2 |H_2|^2 + \lambda_4
|H_1^\dagger H_2|^2 \nonumber \\
&& \;\;\;\; + \; \frac{\lambda_5}{2} \left\{ (H_1^\dagger H_2)^2 + {\rm h.c.} \right\} \;\; ,
\end{eqnarray}
where $\la_1=\la_2=2 \la_3=\la_H$ with $\la_4=\la_5=0$
when compared to our model.
It implies the degenerate mass spectrum for $H_2^{\pm}$ and $H_2^0$.
The mass splitting, nonetheless, arises from the fact $H_2^0$~(DM particle) 
also receives a tiny contribution from $\De_p$ and $G^p_H$ as well as loop contributions. 
In the limit of IHDM the mass splitting is very small, making  
paramount (co)-annihilations, such as $H^{+}_2 H^{-}_2 \to W^+ W^-$, $H^{0}_2 H^{0}_2 \to W^+ W^-$,
$(H^{+}_2 H^{-}_2, H^{0}_2 H^{0}_2) \to Z Z$, $H^{\pm}_2 H^{0}_2 \to W^{\pm} \ga$,
and thus the DM relic density is mostly determined by these channels.

Besides, from Eq.~\eqref{IHDM_pot}, $\la_H$ is seemingly fixed by the Higgs mass, $\la_1= m^2_h/2 v^2 \sim 0.13$, a distinctive feature
of IHDM. 
In G2HDM, however, the value of $\la_H$ can deviate from 0.13 due to the mixing among
$h$, $\de_3$ and $\phi_2$ as shown in Fig.~\ref{fig:laH_mHD}, where the red curve corresponds to the Higgs boson mass of 125 GeV
with $\la_H$ varying from 1.2 to 1.9 as a function of $M_{H\De}$, which controls the scalar mixing.
To simulate G2HDM but still stay close to the IHDM limit, we will treat $\la_H$ as a free parameter 
in the analysis and independent of the SM Higgs mass. 

We note that {{\tt micrOMEGAs}} requires five input parameters for IHDM: SM Higgs mass~(fixed to be 125 GeV), $H^{\pm}_2$ mass,
$H_2^0$ mass~(including both CP-even and odd components which are degenerate in our model), $\la_2~(=\la_H)$ and
$(\la_3+\la_4-\la_5)/2~(\approx \la_H)$. It implies that only two of them are independent in G2HDM, which can be chosen to be $m_{\rm DM}$ and $\lambda_H$.
Strictly speaking, $m_{\rm DM}$ is a function of parameters such as $M_{\Phi\De}$, $M_{H\De}$,
$v_\Phi$, $v_\De$, {\it etc.}, since it is one of the eigenvalues in the mass matrix of Eq.~\eqref{goldstonemassmatrix}.
In this analysis, we stick to the scan range of the parameters displayed in Eq.~\eqref{domain} with slight modification 
as follows
\begin{eqnarray}
\label{domain_1}
0.12 &\leq \, \la_{H} \, \leq & 0.2 \; , \nonumber\\
0.8 &\leq \, M_{\Phi\Delta}/M_{H\Delta} \, \leq & 1.0 \; ,
\end{eqnarray} 
and also demand the mixing~(denoted by $\mathcal{O}_D$) between $H^0_2$ and the other scalars to be less than $1\%$.
In the exact IHDM limit, ${\cal O}_D$ should be 1. 
Besides, the decay time of $H_2^{\pm}$ into $H_2^0$ plus an electron and an electron neutrino 
is required to be much shorter than one second in order not to spoil the Big Bang nucleosynthesis.

\begin{figure}[htbp!]
\centering
\includegraphics[clip,width=0.48\linewidth]{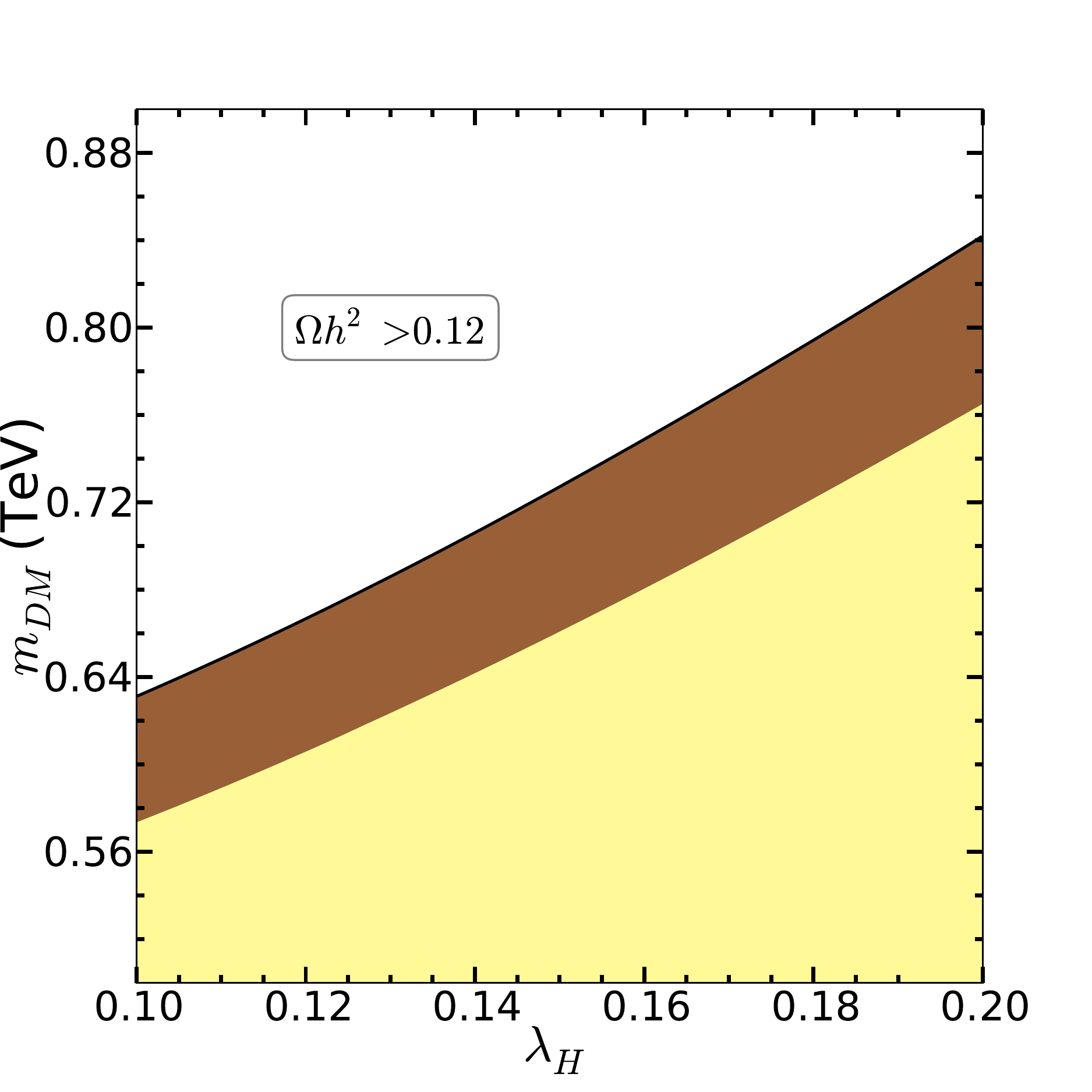}
\includegraphics[clip,width=0.48\linewidth]{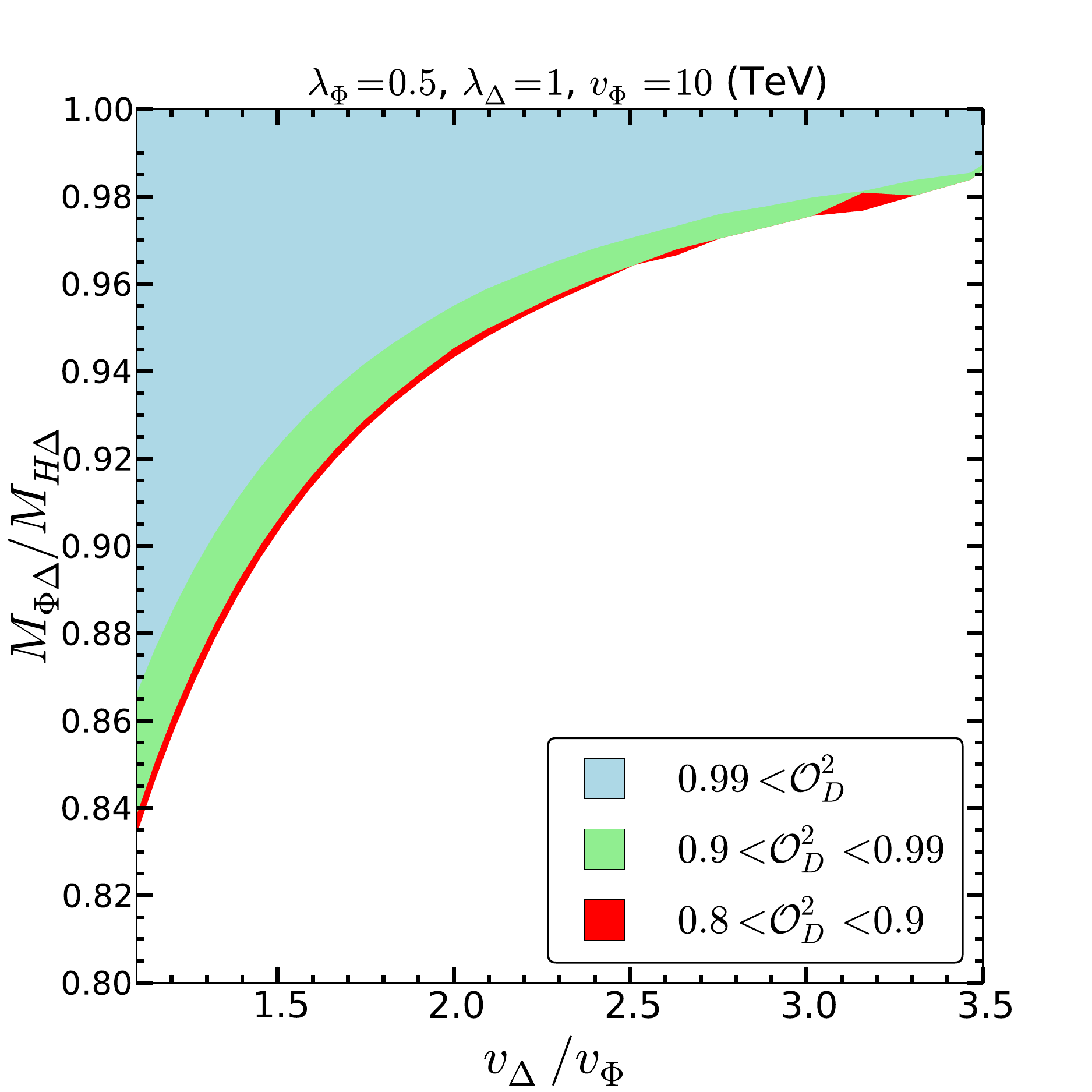}
\caption{ Left: the contour of relic density on ($\lambda_H$, $m_{DM}$) plane. 
The upper brown region is with the relic density between 
0.1 and 0.12 but the lower yellow region is the relic density 
less than 0.1. 
Right: accepted DM mass region projected on ($v_\De/v_\Phi$, $M_{\Phi\De}/M_{H\De}$) plane. 
The red, green and light blue regions present the DM inert Higgs fraction
$0.8<\mathcal{O}^2_D<0.9$, $0.9<\mathcal{O}^2_D<0.99$ and $0.99<\mathcal{O}^2_D$, 
respectively. 
See the text for detail of the analysis.}
\label{fig:DM_vrMr}  
\end{figure}

Our result is presented in Fig.~\ref{fig:DM_vrMr}. 
For a given $\la_H$, there exists an upper bound on the DM mass,
above which the DM density surpasses the observed one, as shown in the left panel of Fig.~\ref{fig:DM_vrMr}.
The brown band in the plot corresponds to the DM relic density of 
$ 0.1 <\Omega h^2 <0.12 $ while the yellow band refers to $ \Omega h^2 < 0.1 $.
In the right panel of Fig.~\ref{fig:DM_vrMr}, we show how well the IHDM limit can be reached
in the parameter space of $M_{\Phi\De}/M_{H\De}$
versus $v_\De/v_\Phi$, where
the red band corresponds to $0.8<\mathcal{O}^2_D<0.9$, the green band refers to $0.9<\mathcal{O}^2_D<0.99$~(green) and
while the light blue area represents $0.99<\mathcal{O}^2_D$.
One can easily see the IHDM limit ~($\mathcal{O}^2_D \sim 1$) can be attained with $M_{\Phi\De} \lesssim M_{H\De}$.
For $M_{\Phi\De} > M_{H\De}$, we have $m_{H^\pm_2} <  m_{H^0_2}$, implying $H_2^0$ can not be DM anymore.
Finally, we would like to point out that the allowed parameter space may increase significantly
once the small $H_2^0$ mixing constraint is removed and other channels comprising $G^p_H$ and $\De_p$ are considered.

It is worthwhile to mention that not only $H_2^0$ but also $\nu^H$~(right-handed neutrino's $SU(2)_H$ partner)
can be the DM candidate in G2HDM, whose stability can be proved similarly 
based on the gauge and Lorentz invariance.
If this is the case, then there exists an intriguing connection between DM phenomenology and neutrino physics.
We, however, will leave this possibility for future work. 

\subsection{EWPT - $\Delta S$, $\Delta T$ and $\Delta U$}

In light of the existence of new scalars and fermions~\footnote{ For additional gauge bosons in the limit of $U(1)_X$ being a global symmetry, the constrain considered above,
$\sin\th_{ZZ^\prime}<10^{-3}$, are actually derived from the electroweak precision data combined with collider bounds. As a consequence, we will not discuss their contributions again.}, one should be concerned about the EWPT, characterized by the oblique parameters $\Delta S$, $\Delta T$
and $\Delta U$~\cite{Peskin:1991sw}.
For scalar contributions, similar to the situation of DM relic density computation, 
a complete calculation in G2HDM can be quite convoluted. The situation, however, becomes much simpler when one goes to the IHDM limit as before, {\it i.e.}, $H_2^0$ is virtually the mass eigenstate
and the inert doublet $H_2$ is the only contribution to $\Delta S$, $\Delta T$
and $\Delta U$, since $\Phi_H$ and $\De_H$ are singlets under the SM gauge group. 
Analytic formulas for the oblique parameters  can be found in Ref.~\cite{Barbieri:2006dq}.
All of them will vanish if the mass of $H_2^\pm$ is equal to that of $H_2^0$ as guaranteed by
the $SU(2)_L$ invariance.

On the other hand, the mass splitting stems from the $H_2^{0*}$ mixing with $G^p_H$ and $\De_p$;
therefore, in the limit of IHDM, the mass splitting is very small, implying a very small deviation from the SM predictions of the oblique parameters. 
We have numerically checked that all points with the correct DM relic abundance studied in the previous section 
have negligible $H_2$ contributions, of order less than $10^{-3}$,
to $\Delta S$, $\Delta T$ and $\Delta U$.

Finally for the heavy fermions, due to the fact they are singlets under $SU(2)_L$,
the corresponding contributions will vanish according to the definition of oblique parameters~\cite{Peskin:1991sw}.  
Our model survives the challenge from the EWPT as long as one is able to
suppress the extra scalar contributions, for instance, resorting to the IHDM limit or having cancellation among different contributions.

  
\section{Conclusions and outlook} \label{section:conclusion}

In this work, we propose a novel framework to embed two Higgs doublets, $H_1$ and $H_2$ into a doublet under
a non-abelian gauge symmetry $SU(2)_H$ and the $SU(2)_H$ doublet is charged under an additional abelian group $U(1)_X$.
To give  masses to additional gauge bosons, we introduce an $SU(2)_H$ scalar triplet and doublet~(singlets under the SM
gauge group). The potential of the two Higgs doublets is as simple as the SM Higgs potential at the cost of additional terms
involving the $SU(2)_H$ triplet and doublet. 
The vev of the triplet triggers the spontaneous symmetry breaking of $SU(2)_L$ by generating the vev for the first
$SU(2)_L$ Higgs doublet, identified as the SM Higgs doublet, while the second Higgs doublet does not obtain a vev and
the neutral component could be the DM candidate, whose stability is guaranteed by the $SU(2)_H$ symmetry and Lorentz invariance.  
Instead of investigating DM phenomenology, we have focused here on Higgs physics and mass spectra of new particles. 

To ensure the model is anomaly-free and SM Yukawa couplings preserve the additional $SU(2)_H \times U(1)_X$ symmetry,
we choose to place the SM right-handed fermions and new heavy right-handed fermions into $SU(2)_H$ doublets while SM $SU(2)_L$
fermion doublets are singlets under $SU(2)_H$. Moreover, the vev of the $SU(2)_H$ scalar doublet can provide a mass to the
new heavy fermions via new Yukawa couplings.   

Different from the left-right symmetric model with a bi-doublet Higgs bosons, in which $W^\pm_R$ carries the electric charge,
the corresponding $W^\prime$ bosons in this model that connects $H_1$ and $H_2$ are electrically neutral since $H_1$ and
$H_2$ have the same SM gauge group charges.
On the other hand, the corresponding $Z^\prime$ actually mixes with the SM $Z$ boson and the mixing are constrained by the
EWPT data as well as collider searches on the $W^\pm$ final state.
$Z^\prime$ itself is also confronted by the direct resonance searches based on dilepton or dijet channels,
limiting the vev of the scalar doublet $\Phi$
to be of order $\mathcal{O}(\text{TeV})$.

By virtue of the mixing between other neutral scalars and the neutral component of $H_1$,
the SM Higgs boson is a linear combination of three scalars.
So the SM Higgs boson tree-level couplings to the SM fermions and gauge bosons are universally rescaled by the mixing angle with respect to
the SM results.
We also check the $h \to \ga\ga$ decay rate normalized to the SM value and it depends not only 
on the mass of $H_2^\pm$ but also other mixing parameters.
As a result, $H_2^\pm$ has to be heavier than $100$ GeV while the $h \to Z \ga$ decay width
is very close to the SM prediction. We also confirm that our model can reproduce the correct DM relic abundance and stays
unscathed from the EWPT data in the limit of IHDM where DM is purely the second neutral Higgs $H^0_2$.
Detailed and systematic study will be pursued elsewhere.

As an outlook, we briefly comment on collider signatures of this model, for which detailed analysis goes beyond
the scope of this work and will be pursued in the future. Due to the $SU(2)_H$ symmetry, searches for heavy
particles are similar to those of SUSY partners of the SM particles with $R$-parity.
In the case of $H^0_2$ being the DM candidate, one can have, for instance, $ u_R  \overline{u_R} \to W^{\prime p} W^{\prime m}$ via $t$-channel exchange of $u^H_R$, followed by $W^{\prime p} \to \overline{u_R} u^H_R \to \overline{u_R} H^0_2 u_L$ and its complex conjugate, leading to 4 jets plus missing transverse energy. Therefore, searches on charginos or gauginos in the context of SUSY may also apply to this model.
Furthermore, this model can also yield mono-jet or mono-photon signatures: $ u_R  \overline{u_R} \to H^0_2 H^0_2 $ plus $\ga$ or $g$ from the initial state radiation. Finally, the recent diboson
excess observed by the ATLAS Collaboration~\cite{Aad:2015owa} may be partially explained by 2 TeV 
$Z^\prime$ decays into $W^+ W^-$ via the $Z^\prime - Z$ mixing.    

Phenomenology of G2HDM is quite rich. In this work we have only touched upon its surface. Many topics like constraints from vacuum stability as well as 
DM and neutrinos physics, collider implications, {\it etc} are worthwhile to be pursued further. We would like to return to some of these issues in the future.

 
 \newpage

\appendix
\section{}\label{section:app} 

From Eqs.~\eqref{vevv} and \eqref{vevphi}, besides the trivial solutions of $v^2=v^2_\Phi=0$ one can deduce the following non-trivial expressions for $v^2$ and $v_\Phi^2$ respectively,
\begin{eqnarray}
v^2 & = & \frac{(2 \lambda_\Phi\lambda_{H\De}-\lambda_{H\Phi}\lambda_{\Phi \De}) v^2_\De + (\lambda_{H\Phi}M_{\Phi \De}-2\lambda_\Phi M_{H\De}) v_\De + 2 (2 \lambda_\Phi \mu^2_H - \lambda_{H \Phi} \mu^2_\Phi )} {\lambda _{H\Phi }^2-4 \lambda _H \lambda _{\Phi }} \; , \\
v^2_\Phi & = & \frac{(2 \lambda_{H}\lambda_{\Phi \De}-\lambda_{H\Phi}\lambda_{H \De}) v^2_\De + (\lambda_{H\Phi}M_{H \De}-2\lambda_H M_{\Phi\De}) v_\De + 2 (2 \lambda_H \mu^2_\Phi - \lambda_{H \Phi} \mu^2_H )}{\lambda _{{H\Phi }}^2-4 \lambda _H \lambda _{\Phi }} \; .
\end{eqnarray}
Substituting the above expressions for $v^2$ and $v_\Phi^2$ into Eq.~\eqref{vevdelta} leads to the following
cubic equation for $v_\De$:
\begin{equation}
v_\De^3 + a_2 v_\De^2 + a_1 v_\De + a_0 = 0 \; ,
\label{cubiceq}
\end{equation}
where $a_2=C_2/C_3$, $a_1=C_1/C_3$ and $a_0=C_0/C_3$ with
\begin{eqnarray}
C_0 & = & 
2 \left(\lambda _{{H\Phi }} M_{\Phi \Delta }-2 \lambda _{\Phi }M_{{H\Delta }}\right) \mu _H^2 +2  \left(\lambda _{{H\Phi }}M_{{H\Delta }} -2 \lambda _H M_{\Phi \Delta }\right) \mu _{\Phi }^2 \; ,\\
C_1 & = & 2 \left[  
 2 \left( 2 \lambda _{{H\Delta }} \lambda _{\Phi }- \lambda _{{H\Phi }} \lambda _{\Phi \Delta } \right) \mu _H^2  + 2  \left( 2 \lambda _H \lambda _{\Phi \Delta } -
\lambda _{{H\Delta }} \lambda _{{H\Phi }} \right)\mu _{\Phi }^2\right. \nonumber \\
&& \;\;\; + 2 \left(4  \lambda _H \lambda _{\Phi }  -  \lambda _{{H\Phi }}^2 \right) \mu _{\Delta }^2 + \left. \lambda _H M_{\Phi \Delta }^2-\lambda _{{H\Phi }} M_{{H\Delta }} M_{\Phi \Delta }+\lambda _{\Phi }M_{{H\Delta }}^2 \right] \; ,
\\
C_2 & = & 3 \left[  \left(\lambda _{{H\Delta }} \lambda _{{H\Phi }}   -2 \lambda _H \lambda _{\Phi \Delta }\right)M_{\Phi \Delta } +\left(\lambda _{{H\Phi }} \lambda _{\Phi \Delta }-2 \lambda _{{H\Delta }} \lambda _{\Phi }\right) M_{{H\Delta }} \right] \; ,\\
C_3 & = & 4 \left[ \lambda _H \left(\lambda _{\Phi \Delta }^2-4 \lambda _{\Delta } \lambda _{\Phi }\right)-\lambda _{{H\Delta }} \lambda _{{H\Phi }} \lambda _{\Phi \Delta }+\lambda _{{H\Delta }}^2 \lambda _{\Phi }+\lambda _{\Delta } \lambda _{{H\Phi }}^2\right] \; .
\end{eqnarray}
The three roots of cubic equation like Eq.~\eqref{cubiceq} are well-known since the middle of 16$^{\rm th}$ century 
\begin{eqnarray}
v_{\De \, 1} & = & -\frac{1}{3} a_2 + \left( S + T \right) \; ,\\
v_{\De \, 2} & = & -\frac{1}{3} a_2 - \frac{1}{2} \left( S + T \right) + \frac{1}{2} i {\sqrt 3} \left( S - T \right) \; ,\\
v_{\De \, 3} & = & -\frac{1}{3} a_2 - \frac{1}{2} \left( S + T \right) - \frac{1}{2} i {\sqrt 3} \left( S - T \right) \; ,
\label{cubicroots}
\end{eqnarray}
where
\begin{eqnarray}
S & \equiv & \, ^3\sqrt{R + \sqrt D} \;, \\
T & \equiv & \, ^3\sqrt{R - \sqrt D} \;, \\
D & \equiv & \, Q^3 + R^2 \; ,
\end{eqnarray}
with
\begin{eqnarray}
Q & \equiv & \frac{3 a_1 - a_2^2}{9} \; , \\
R & \equiv & \frac{9 a_1 a_2 - 27 a_0 - 2 a_2^3}{54} \; .
\end{eqnarray}


\section{}\label{section:app-HDecays}

Below we summarize the results for the decay width of SM Higgs to $\gamma\gamma$ and $\gamma Z$~\cite{Gunion:1989we,Djouadi:2005gi,Djouadi:2005gj,Chen:2013vi},
including the mixing among $h$, $\de_3$ and $\phi_2$ characterized by the orthogonal matrix $O$, i.e., $(h, \de_3, \phi_2)^T = O \cdot (h_1, h_2, h_3)^T$. 
In general one should include the mixing effects among $Z$ and $Z'$ (and perhaps $Z''$) as well. As shown in Section IV,
these mixings are constrained to be quite small and we will ignore them here.

\begin{itemize}

\item 
Taking into account $H^{\pm}_2$ contributions, the partial width of $h_1 \to \ga\ga$ is
\begin{eqnarray}
\Gamma \, (h_1\to \gamma\gamma) = \frac{G_{F}\, \alpha^{2}\,m_{h_1}^{3}  O_{11}^2  }
{128\,\sqrt{2}\,\pi^{3}} 
\left| \mathcal{C}_h
\frac{  \la_H v^2}{m_{H_2^\pm}^2} A_0^{\ga\ga}(\tau_{H_2^\pm})+
A^{\ga\ga}_1(\tau_W) + 
\sum_{f} N_{c} Q_f^2 A_{1/2}^{\ga\ga}(\tau_f) 
\right|^2 , \nn \\
\label{eq:hgagaH2}
\end{eqnarray}
with
\begin{align}
\mathcal{C}_h = 1 - \frac{O_{21}}{O_{11}} \frac{ 2 \la_{H\De} v_{\De} + M_{H\De} }{ 4 \la_H v}
+ \frac{O_{31}}{O_{11}} \frac{ \la_{H\Phi} v_{\Phi} }{ 2 \la_H v} \; .
\end{align}
The form factors for spins $0$, $\frac{1}{2}$ and 1 particles are given by
\begin{eqnarray}
A_{0}^{\ga\ga}(\tau) & =  & - [\tau -f(\tau)]\, \tau^{-2} \, ,\nonumber \\
A_{1/2}^{\ga\ga}(\tau) & = & 2 [\tau +(\tau -1)f(\tau)]\, \tau^{-2}  \, , \nonumber \\   
A_1^{\ga\ga}(\tau) & = & - [2\tau^2 +3\tau+3(2\tau -1)f(\tau)]\, \tau^{-2} \, ,
\label{eq:A0+Af+Aw}
\end{eqnarray}
with the function $f(\tau)$ defined by
\begin{eqnarray}
f(\tau)=\left\{
\begin{array}{ll}  \displaystyle
\arcsin^2\sqrt{\tau} & {\rm , \;\; for} \; \tau\leq 1 \; ; \\
\displaystyle -\frac{1}{4}\left[ \log\frac{1+\sqrt{1-\tau^{-1}}}
{1-\sqrt{1-\tau^{-1}}}-i\pi \right]^2 \hspace{0.5cm} & {\rm , \; \; for} \; \tau>1 \; .
\end{array} \right.
\label{eq:ftau}
\end{eqnarray}
The parameters $\tau_i= m_{h_1}^2/4m_i^2$ with $i=H_2^\pm,f,W^\pm$ are related to the
corresponding masses of the heavy particles in the loops.
%


\item  
Including the $H^\pm_2$ contribution, we have
\begin{align}
\Gamma (h_1\to Z\gamma ) &= \frac{G^2_{F} m_W^2\, \alpha\,m_{h_1}^{3} O_{11}^2} 
{64\,\pi^{4}} \left( 1-\frac{m_Z^2}{m_{h_1}^2} \right)^3\times \nn\\&\left|
\left.  - \mathcal{C}_h \frac{ \la_H v^2 } {  m_{H_2^\pm}^2 }v_{H_2^\pm}\,A^{ Z \ga}_0 (\tau_{H_2^\pm},\lambda_{H_2^\pm}) +
A^{Z \ga}_1(\tau_W,\lambda_W)  \right. +   
\sum_{f} N_{c} \frac{Q_f \hat{v}_f}{c_W} A_{1/2}^{Z \ga}(\tau_f,\lambda_f)  
\right|^2 ,
\label{eq:hzgaH2}
\end{align}
with
$v_{H_2^\pm}  = (2c_W^2-1)/c_W$,
$\hat{v}_f  =  2I_f^3-4 Q_f s_W^2$,
$\tau_i= 4m_i^2/m_{h_1}^2$ and $\lambda_i = 4m_i^2 /m_Z^2$. 
The loop functions are
\begin{eqnarray}
A^{Z \ga}_0 (\tau_{H^\pm},\lambda_{H^\pm}) & = &
I_1 (\tau_{H^\pm},\lambda_{H^\pm}) \; , \nonumber \\
A_{1/2}^{Z \ga} (\tau,\lambda) & = & \left[I_1(\tau,\lambda) - I_2(\tau,\lambda)
\right]  \; ,  \\
A^{Z \ga}_1 (\tau,\lambda) & = & c_W \left\{ 4\left(3-\frac{s_W^2}{c_W^2} \right)
I_2(\tau,\lambda) + \left[ \left(1+\frac{2}{\tau}\right) \frac{s_W^2}{c_W^2}
- \left(5+\frac{2}{\tau} \right) \right] I_1(\tau,\lambda) \right\} \, , \nn
\label{eq:hzgaform}
\end{eqnarray}
where $I_1$ and $I_2$ 
are defined as
\begin{eqnarray}
I_1(\tau,\lambda) & = & \frac{\tau\lambda}{2(\tau-\lambda)}
+ \frac{\tau^2\lambda^2}{2(\tau-\lambda)^2} \left[ f(\tau^{-1})-f(\lambda^{-1}) 
\right] + \frac{\tau^2\lambda}{(\tau-\lambda)^2} \left[ g(\tau ^{-1}) - 
g(\lambda^{-1}) \right] \, , \nn \\
I_2(\tau,\lambda) & = & - \frac{\tau\lambda}{2(\tau-\lambda)}\left[ f(\tau
^{-1})- f(\lambda^{-1}) \right] \, ,
\end{eqnarray}
with the function $f(\tau)$ defined in Eq.~\eqref{eq:ftau} and
the function $g(\tau)$ can be expressed as
\begin{equation}
g(\tau) = \left\{ \begin{array}{ll}
\displaystyle \sqrt{\tau^{-1}-1} \arcsin \sqrt{\tau} & {\rm  , \; \; for} \; \tau \ge 1 \, ; \\
\displaystyle \frac{\sqrt{1-\tau^{-1}}}{2} \left[ \log \frac{1+\sqrt{1-\tau
^{-1}}}{1-\sqrt{1-\tau^{-1}}} - i\pi \right] & {\rm  , \; \; for} \; \tau  < 1 \, .
\end{array} \right.
\label{eq:gtau}
\end{equation}

\end{itemize}
The corresponding SM rates $\Gamma_{\rm SM} (h_1 \to \ga\ga)$ and $\Gamma_{\rm SM} (h_1 \to Z\ga)$ can be obtained by omitting the $H^{\pm}_2$ contribution and
setting $O_{11}=1$ in Eqs.~\eqref{eq:hgagaH2} and \eqref{eq:hzgaH2}.

\newpage 

\section*{Acknowledgments}
The authors would like to thank A. Arhrib, F. Deppisch, M. Fukugita, J. Harz and T. Yanagida for useful discussions.
WCH is grateful for the hospitality of IOP Academia Sinica and NCTS in Taiwan
and HEP group at Northwestern University where part of this work was carried out.
TCY is grateful for the hospitality of IPMU where this project was completed.
This work is supported in part by the Ministry of Science and Technology (MoST) of Taiwan under
grant numbers 101-2112-M-001-005-MY3 and 104-2112-M-001-001-MY3 (TCY), 
the London Centre for Terauniverse Studies (LCTS) 
using funding from the European Research Council via the Advanced Investigator Grant 267352 (WCH),
DGF Grant No. PA 803/10-1 (WCH), 
and the World Premier International Research Center Initiative (WPI), MEXT, Japan (YST).

\newpage

\bibliographystyle{h-physrev}
\bibliography{SU2_H}

\end{document}